\documentclass[aip,amsmath,amssymb,preprint,floatfix]{revtex4-1}
 
\usepackage{graphicx}
\usepackage{epsfig}
\usepackage{color}
\usepackage{dcolumn}
\usepackage{bm}

\usepackage[utf8]{inputenc}
\usepackage[T1]{fontenc}
\usepackage{mathptmx}
\usepackage{caption}

\begin{document}

\title{Multicaloric effects in Metamagnetic Heusler Ni-Mn-In under uniaxial stress and magnetic field.}

\author{Adri\`a Gr\`acia-Condal}
\affiliation{Departament de F\'{\i}sica de la Mat\`eria Condensada, Facultat de F\'{\i}sica, Mart\'{\i} i Franqu\`es 1, Universitat de Barcelona, 08028 Barcelona, Catalonia.}

\author{Tino Gottschall}
\affiliation{Dresden High Magnetic Field Laboratory (HLD-EMFL), Helmholtz-Zentrum, Dresden-Rossendorf, 01328 Dresden, Germany}
\affiliation{Institut f\"ur Materialwissenschaft, Technical University of Darmstadt,  
64287 Darmstadt, Germany}


\author{Lukas Pfeuffer}
\affiliation{Institut f\"ur Materialwissenschaft, Technical University of Darmstadt,  
64287 Darmstadt, Germany}

\author{Oliver Gutfleisch}
\affiliation{Institut f\"ur Materialwissenschaft, Technical University of Darmstadt,  
64287 Darmstadt, Germany}

\author{Antoni Planes}
\affiliation{Departament de F\'{\i}sica de la Mat\`eria Condensada, Facultat de F\'{\i}sica, Mart\'{\i} i Franqu\`es 1, Universitat de Barcelona, 08028 Barcelona, Catalonia.}

\author{Llu\'{\i}s Ma\~nosa}
\affiliation{Departament de F\'{\i}sica de la Mat\`eria Condensada, Facultat de F\'{\i}sica, Mart\'{\i} i Franqu\`es 1, Universitat de Barcelona, 08028 Barcelona, Catalonia.}

\date{\today}


\begin{abstract}

The world's growing hunger for artificial cold on the one hand, and the ever more stringent climate targets on the other, pose an enormous challenge to mankind. Novel, efficient and environmentally friendly refrigeration technologies based on solid-state refrigerants can offer a way out of the problems arising from climate-damaging substances used in conventional vapor-compressors. Multicaloric materials stand out because of their large temperature changes which can be induced by the application of different external stimuli such as a magnetic, electric, or a mechanical field. Despite the high potential for applications and the interesting physics of this group of materials, only few studies focus on their investigation by direct methods. In this paper, we report on the advanced characterization of all relevant physical quantities that determine the multicaloric effect of a Ni-Mn-In Heusler compound. We have used a purpose-designed calorimeter to determine the isothermal entropy and adiabatic temperature changes resulting from the combined action of magnetic field and uniaxial stress on this metamagnetic shape-memory alloy. From these results, we can conclude that the multicaloric response of this alloy by appropriate changes of uniaxial stress and magnetic field largely outperforms the caloric response of the alloy when subjected to only a single stimulus. We anticipate that our findings can be applied to other multicaloric materials, thus inspiring the development of refrigeration devices based on the multicaloric effect.


\end{abstract}

\maketitle

\section{Introduction}

Our society is facing an increasing demand for refrigeration \cite{Shah2017} going hand in hand with a pronounced increase in the energy spent for various cooling applications. Furthermore, conventional refrigeration systems use fluids with a strong global warming potential \cite{climatepolicies} (their effect per unit mass is about thousand times larger than carbon dioxide).  It is therefore urgent to develop technologies which are both energy efficient and respectful with the environment. Solid-state cooling, based on the giant caloric effects exhibited by a variety of materials undergoing ferroic phase transitions, are considered as the best alternatives to replace present refrigerators that use harmful fluids \cite{Manosa2013,Moya2014}. For the last two decades,  a worldwide intensive research  has been devoted to the study of  magnetocaloric \cite{Pecharsky1997}, electrocaloric  \cite{Mischenko2006}, and  mechanocaloric (which include elastocaloric and barocaloric) materials \cite{Bonnot2008,Manosa2010,Cazorla2019}, and very recently the study has been extended to multicaloric materials for which good prospects are envisaged \cite{Liu2012,SternTaulats2018,Gottschall2018}. 

A broad variety of materials exhibiting  large isothermal entropy  and adiabatic temperature changes, induced by the application or removal of different fields (magnetic, electric, mechanical)  have been discovered up to now  \cite{Gschneidner2005,Smith2012,Franco2018,Valant2012,Manosa2017}. This includes recent reports on plastic crystals \cite{Li2019,Lloveras2019} for which the pressure-induced entropy changes  compare to the values given by standard commercial fluid refrigerants (although lower pressures are required for the latter). Elastocaloric materials are known to exhibit the largest values for the adiabatic temperature changes among all caloric materials, and very recently reversible values larger than 30 K have been reported (at readily accessible stresses) for the elastocaloric effect of shape memory alloys formed by all 3{\it d} elements \cite{Cong2019}.

In spite of the significance of the achievements up to now, there are a series of bottlenecks, mostly related to the first-order character of the phase transitions that are at the origin of the giant caloric effects. On the one hand, the required fields to achieve   giant values are in general still too large, and on the other hand, the hysteresis associated with the phase transitions reduces the caloric efficiency and compromises the reversibility of the caloric effect. It has been proposed that the combination of more than one external field may help in overcoming some of the mentioned limitations \cite{Liu2012,SternTaulats2018,Czernuszewicz2018}.

Materials with significant coupling between degrees of freedom have a cross-response to different external stimuli. In those materials, entropy and temperature changes can be driven by either a single stimulus (single caloric effect) or by multiple stimuli (multicaloric effect), which can be applied/removed either simultaneously or sequentially \cite{Planes2014,SternTaulats2018}. The study of multicaloric effects and materials is a quite novel research field, but very interesting results have already been achieved \cite{Gong2015,Liu2016,Liang2019}. It has been shown that a suitable combination of magnetic field and hydrostatic pressure enables a drastic reduction of the magnetic-field-effective hysteresis in materials with magnetostructural transformations \cite{Liu2012,SternTaulats2017,GraciaCondal2018b}. In a recent work, we have undertaken a different strategy to show that in a metamagnetic Ni-Mn-In shape-memory alloy the combination of uniaxial stress and magnetic field enables designing a multicaloric cycle which now takes advantage of the thermal hysteresis associated with the martensitic transition (which is termed "hysteresis positive" approach) \cite{Gottschall2018}. However, the study of multicaloric effects is still challenging. Although the thermodynamic framework is well established \cite{Planes2014}, experimental studies are very scarce \cite{SternTaulats2017,Czernuszewicz2018,Liang2019}, mostly because the obtention of the physical quantities describing these multicaloric effects  typically requires the use of  non-commercial advanced characterization systems \cite{Gottschall2020}.

\begin{figure*}[ht]
\epsfig{file=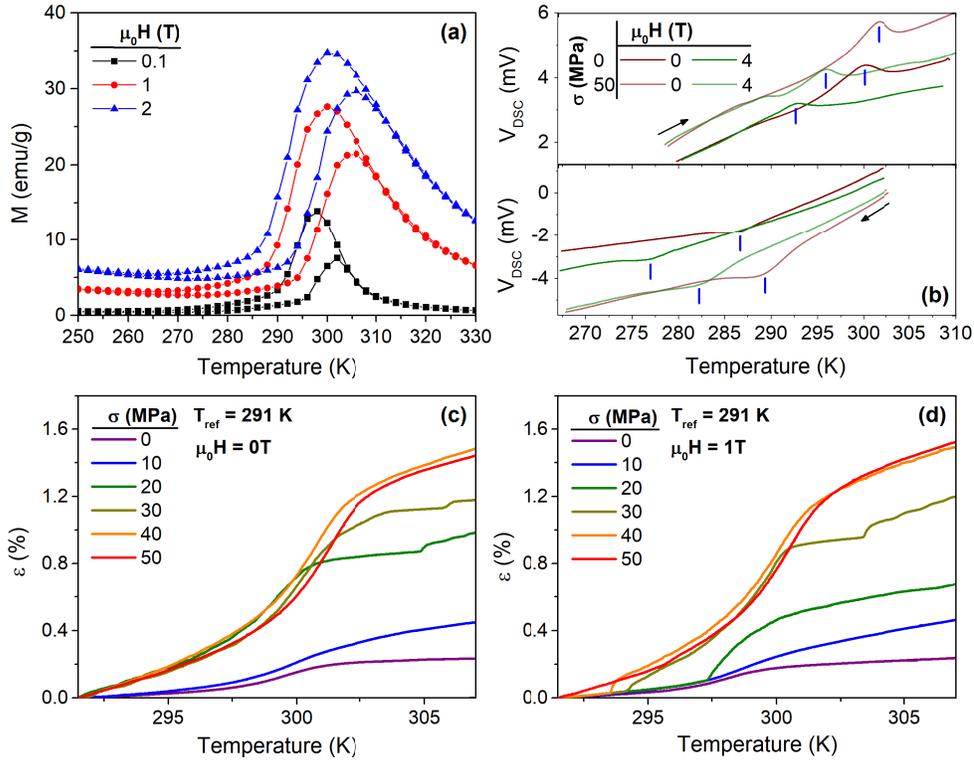,width=16cm,clip=}
\caption{(a) Thermomagnetization curves at selected values of applied magnetic field in the absence of applied stress. (b) Calorimetric signal as a function of temperature for applied stresses of 0 and 50 MPa, and applied magnetic fields of 0 and 4 T. The colour code is indicated in the figure. Upper panel corresponds to heating runs and lower panel corresponds to cooling runs. Vertical blue lines indicate the position of the calorimetric peak.  Panels (c) and (d) show the temperature dependence of strain in the absence of magnetic field (c) and under  1 T applied magnetic field (d), for applied uniaxial stresses of 0, 10, 20, 30, 40 and 50 MPa.}
\label{fig1}
\end{figure*}

In this work, we have used a purpose-built calorimeter that works under the application of uniaxial stress and magnetic field, to  study  the multicaloric response in terms of isothermal entropy and adiabatic temperature changes of a prototypical metamagnetic shape memory alloy subjected to the combined application of magnetic field and uniaxial stress.  We have selected a Ni-Mn-In alloy with a martensitic transition temperature close to the austenitic Curie temperature. The proximity between martensitic and magnetic transitions results in a pronounced coupling between magnetism and structure so that the application of magnetic field has a strong influence on the martensitic transition. Our results evidence the advantages of the multicaloric effect in comparison with the single caloric (magnetocaloric and elastocaloric) ones.

\section{Results}

\begin{figure*}[ht]
\epsfig{file=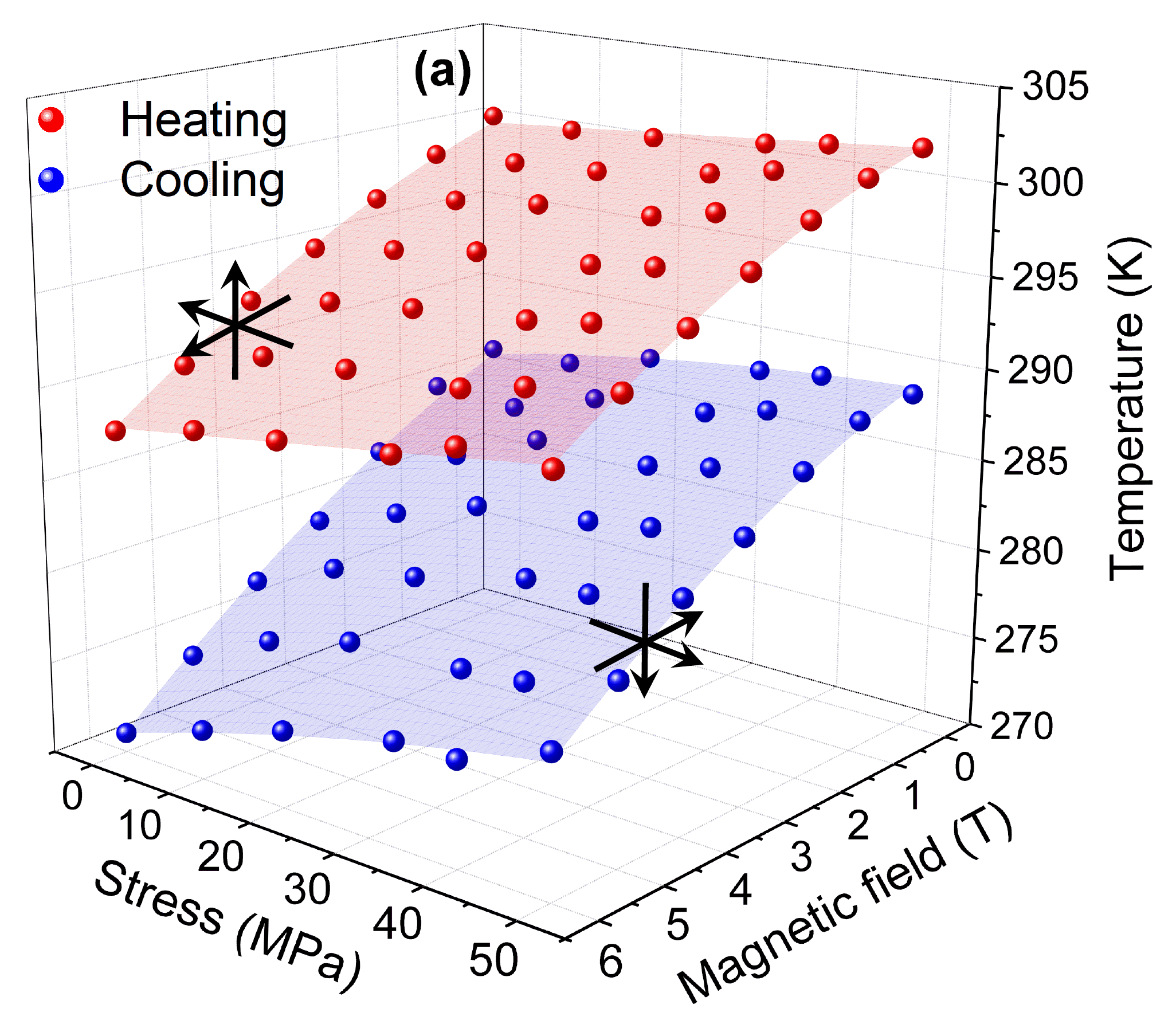,width=6cm,clip=}
\epsfig{file=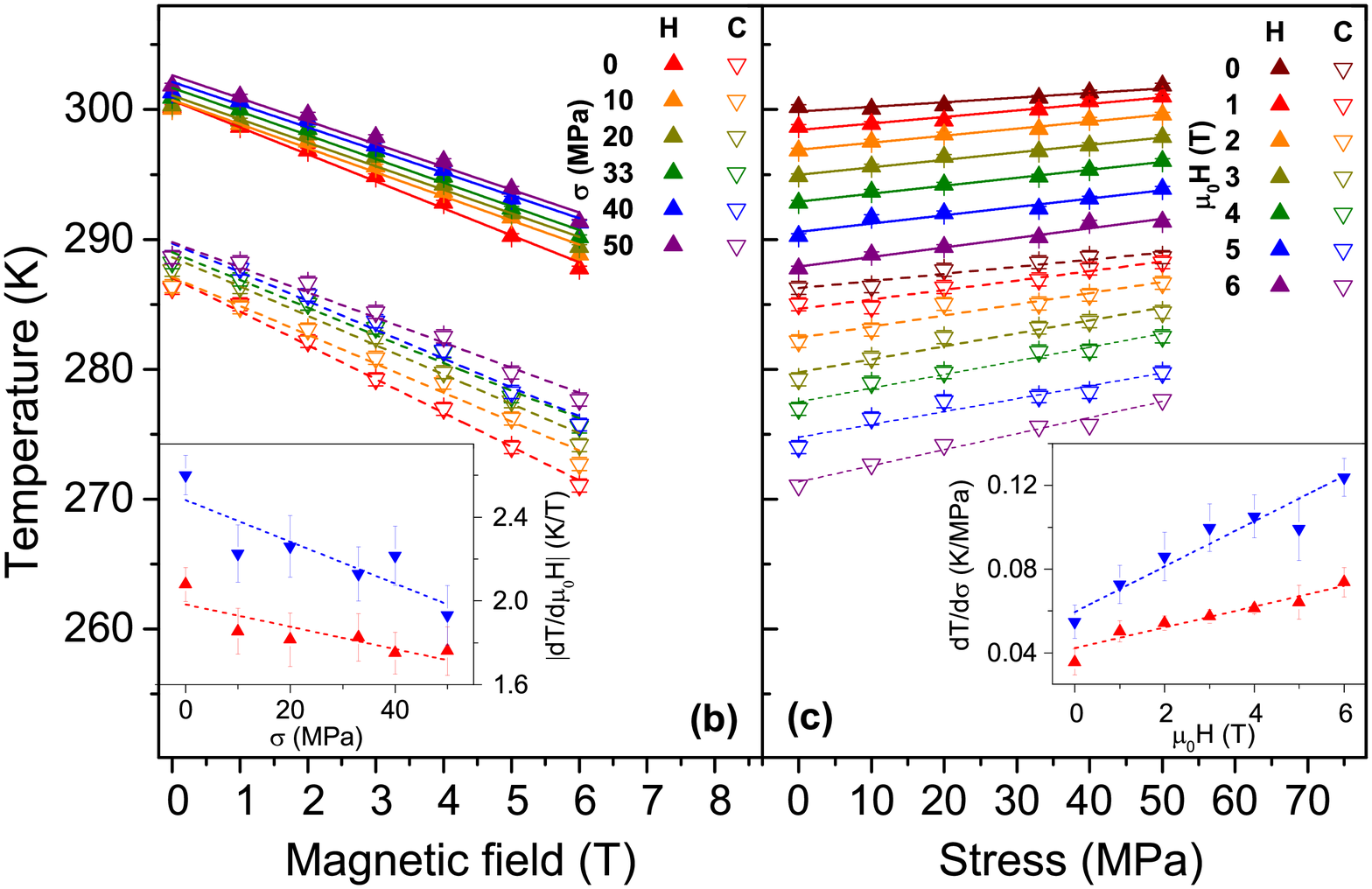,width=8cm,clip=}
\caption{(a)  Transition temperatures as a function of magnetic field and uniaxial stress. Solid symbols indicate experimental data and planes correspond to the best fits to the experimental values. Upper red plane corresponds to the reverse martensitic transition and lower blue plane to the forward martensitic transition. The arrows in each plane indicate the changes in temperature, magnetic field and stress to cross the plane (i.e. to induce the forward and reverse transitions, respectively). (b) Transition temperatures as a function of magnetic field (at constant uniaxial stress) and (c) transition temperatures as a function of uniaxial stress (at constant magnetic field). Solid symbols correspond to the reverse martensitic transition and open symbols to the forward martensitic transition. The inset in figure (b) shows the stress dependence of the slope of the transition temperature {\it vs} magnetic-field lines, and the inset in figure (c) shows the magnetic-field dependence of the slope of the transition temperature {\it vs} stress lines. Lines are linear fits to the data.}
\label{fig2}
\end{figure*}

Thermomagnetization curves were recorded as a function of temperature for selected values of magnetic field ($H$). Our results  (Fig. 1(a)) conform to the reported behaviour for metamagnetic Ni-Mn-In Heusler alloys \cite{Krenke2006,Planes2009}. On cooling, magnetization increases due to the onset of the ferromagnetic order in the austenite ($ T_c \sim$ 303 K), and upon further cooling magnetization sharply decreases at the martensitic transition, as a consequence of the short range antiferromagnetic interactions in the martensitic phase of Ni-Mn-In alloys \cite{Aksoy2009}. 

Calorimetric measurements were conducted at 0.6 and 1 Kmin$^{-1}$ for heating and cooling runs, respectively, for selected values of applied (constant) magnetic field  in the range 0 - 6 T, and applied (constant) uniaxial load in the range 0 - 1 kN. Changes in the cross-section of the specimen are expected to be negligibly small and stresses ($\sigma$) correspond to the ratio between applied load and the unstressed specimen's cross-section measured at room temperature. Examples of the measured calorimetric curves are shown in Fig. 1(b)  for 0 and 4 T magnetic field, and 0 and 50 MPa uniaxial stress. On cooling, the exothermal peak corresponds to the forward martensitic transition while the endothermal peak on heating corresponds to the reverse martensitic transition.

%

From the measurement of the specimen's length $l(T, \sigma, H)$, we have computed the temperature dependence of the strain $\varepsilon$, as $\varepsilon$ = 100 $\times \frac{l(T,H,\sigma)-l_0}{l_0}$ where $l_0$ is the measured length of the sample in the absence of stress and of magnetic field at a temperature $T_{ref}$ = 291 K. Examples for heating runs are shown in Figs. 1(c) and  1(d) for selected values of magnetic field and uniaxial stress. The transition strain  increases with increasing uniaxial stress, as a result of the increase in the percentage of favourably oriented martensitic variants \cite{Karaca2009}. It is worth noticing the good correlation for the transition region from the two sets of measurements, which indicates that both latent heat release (absorption) and strain are proportional to the transformed fraction. 

From both, calorimetric and strain data, it is apparent that the application of a magnetic field shifts the martensitic transition to lower temperatures while uniaxial stress shifts the transition to higher temperatures. By identifying the transition temperature from the peak position of the calorimetric curves on heating and cooling, it is possible to determine the phase diagram in the $H - \sigma$  coordinate space, which is shown in Fig. 2(a). The red plane corresponds to the reverse martensitic transition which occurs either by increasing temperature, increasing magnetic field or decreasing stress (as indicated by arrows in the figure). Well above that plane the sample is in austenite \cite{footnote1}. The blue plane corresponds to the forward martensitic transition induced either by decreasing temperature, decreasing magnetic field or increasing stress (as indicated by arrows in the figure). Well below that plane the sample is in martensite \cite{footnote1}. The region in between the two planes accounts for the hysteresis where the state of the sample is history dependent. The projections  on the $T$-$H$ and $T$-$\sigma$  planes are shown in Figs. 2(b) and  2(c), respectively. For all values of applied stress, both the forward and reverse martensitic transition temperatures linearly decrease with increasing field, with a slope in the absence of applied stress that compares well with typical data for this kind of alloys \cite{Planes2009}. Also, for all values of applied magnetic field, transition temperatures linearly increase with increasing stress. In this case, the slope of the transition lines in the absence of magnetic field is lower but comparable to the values reported  for similar alloys \cite{Karaca2009}. The slope of these transition lines has been found to depend on the application of the secondary field, as illustrated in the inset of Fig. 2(b), which shows a decrease in $|dT/d\mu_0 H|$ with increasing stress and in the inset of Fig. 2(c) which shows an increase in  $dT/d\sigma$ with increasing magnetic field. There are very few studies on the caloric response of materials subjected to more than one external stimulus, and most of them refer to the combined effect of hydrostatic pressure and magnetic field \cite{SternTaulats2017,Liang2019,Hao2020}. The effect of uniaxial stress and magnetic field was studied for the metamagnetic transition in Fe-Rh \cite{GraciaCondal2018b}. For that compound, it was found that the application of a secondary field does not affect the values for $|dT/d\mu_0 H|$ and $dT/d\sigma$.

\section{Data analysis and Discussion}

\subsection{Transition entropy change and entropy curves}

\begin{figure*}
\epsfig{file=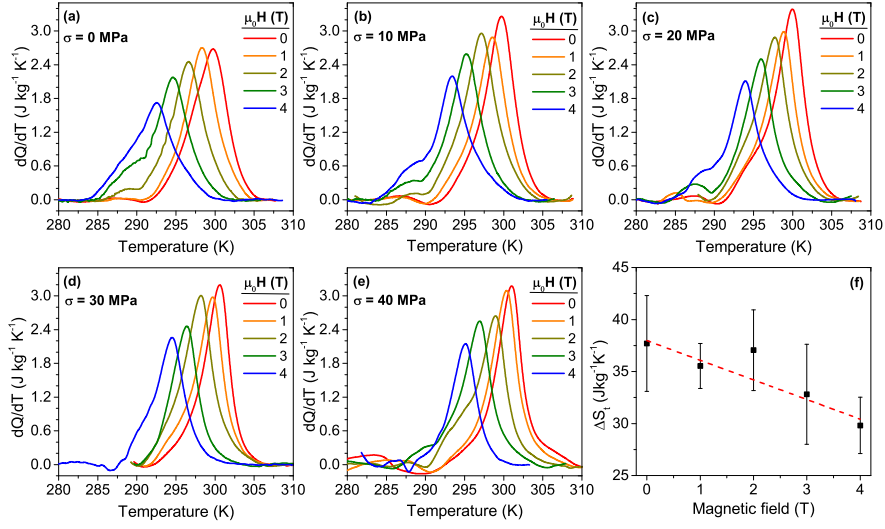,width=12cm,clip=}
\caption{(a)-(e) Base-line corrected thermal curves corresponding to heating runs for applied magnetic fields of 0, 1, 2, 3, and 4 T and applied uniaxial stresses of 0, 10, 20, 30, and 40 MPa. (f) Transition entropy change (computed from heating runs) as a function of magnetic field. The error bars include data obtained for different values of applied uniaxial stress in the range 0-40 MPa. The dashed line is a linear fit to the data.}
\label{fig3}
\end{figure*}

The complexity of the experimental set-up allowing the application of uniaxial load makes our calorimeter less accurate than a conventional DSC. As a result, calorimetric signals are affected by a poorer base-line, particularly  on cooling runs and at high magnetic fields. For this reason, we will restrict the following analysis to the heating protocol and magnetic fields up to 4 T. 

The transition entropy change ($\Delta S_t$) can be computed by suitable integration of the base-line corrected calorimetric curves (Fig. 3(a-e)) as 

\begin{equation}
\Delta S_t (\sigma , H) = \int_{T_1}^{T_2} \frac{1}{T}
\frac{d{Q}(T, \sigma, H)}{dT} dT,
\end{equation}

\noindent where $\frac{dQ(T,\sigma , H)}{dT} = \frac{\dot{Q}(T,\sigma , H)}{\dot{T}}$, with $\dot{Q}(T,\sigma, H)$ being the heat flux measured while scanning temperature at constant stress and magnetic field, and $\dot{T}$ is the heating  rate. $T_1$ and $T_2$ are, respectively, freely chosen temperatures below and above the phase transition region. 

For the range of applied stresses, we have not found any systematic dependence of $\Delta S_t$ with applied stress, but conversely $\Delta S_t$  decreases with increasing magnetic field, as illustrated in Fig. 3(f) which shows $\Delta S_t$ as a function of magnetic field. For each applied field, $\Delta S_t$ values obtained for different applied stresses are within the experimental error (indicated by the error bars). The observed decrease in $\Delta S_t$  with increasing magnetic field is a consequence of the increase in the magnetic contribution to the total entropy, which opposes the phonon contribution being larger (in absolute value) than its magnetic counterpart, and to a good approximation, magnetic field independent  \cite{Kihara2014,Gottschall2016}.



The entropy curves ($S(T, \sigma, H)$) referenced to a value at a specific temperature $T_0$ (chosen below the phase transition region) can be computed \cite{Manosa2017} by combining calorimetric curves (recorded at selected values of stress and magnetic field) with specific heat ($C$) data . In these computations it is common to take $C$ to be independent of magnetic field and stress. While the stress independence of $C$ is still a good approximation in our case, the proximity of the martensitic transition to the austenitic Curie temperature in our sample makes $C$ to be dependent on the applied magnetic field. For this reason, we have measured the temperature dependence of $C$ at selected magnetic fields. Results for the sample under study (Ni$_{50}$Mn$_{35.5}$In$_{14.5}$) are shown in Fig. 4(a), which have been measured using a bespoke  universal calorimeter that operates up to a temperature of 310 K and under an external magnetic field \cite{Stotter 2019}. From the figure, three different regions with distinct behaviour for $C$ vs $T$ are clearly identified. At low temperatures, below the martensitic transition there is no dependence of the specific heat of the martensite ($C_M$) with the magnetic field, and a linear temperature dependence can be assumed (black line in the inset of Fig. 4(a)). Within the transition region (around room temperature), the latent heat of the phase transition results in an apparent peak in the specific heat. This peak shifts to lower temperature under a 2 T magnetic field, which is in good agreement with the shift observed in DSC data (shown in Fig. 3). Above the martensitic transition, a small peak (centered at 303 K) associated with the Curie point of the austenite is clearly visible for the 0 T curve. This peak is smoothed when a 2 T magnetic field is applied. Because our bespoke calorimeter is limited at high temperatures, it is not possible to accurately determine the specific heat in the austenitic phase ($C^A$). For this reason, we have used a second (commercial) calorimeter (PPMS from Quantum Design) to measure $C$ over a broader range of temperatures and magnetic fields. In order to separate the contributions from the latent heat and the Curie point we have used a second sample (Ni$_{50}$Mn$_{34}$In$_{16}$) for which the martensitic transition takes place at a temperature (180 K) well below its Curie point. It is not expected that small differences in composition may affect the value of $C^A$ in the paramagnetic austenitic phase. Results from these measurements are summarized in Fig. 4(b). As shown in the inset of Fig. 4(b), above the Curie temperature, $C^A$ is almost independent of temperature but it shows a small dependence in magnetic field.

\begin{figure*}
\epsfig{file=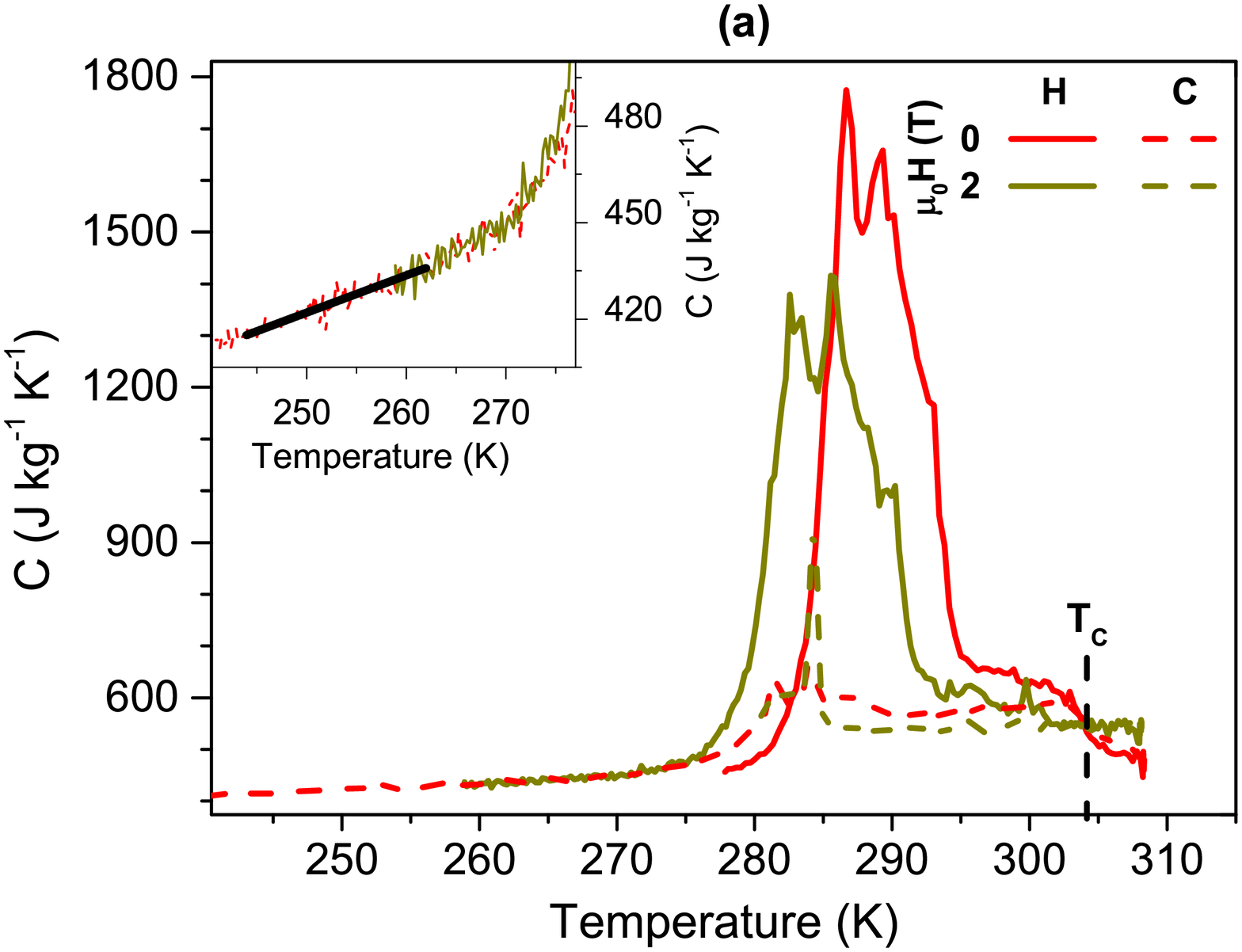,width=8cm,clip=}
\epsfig{file=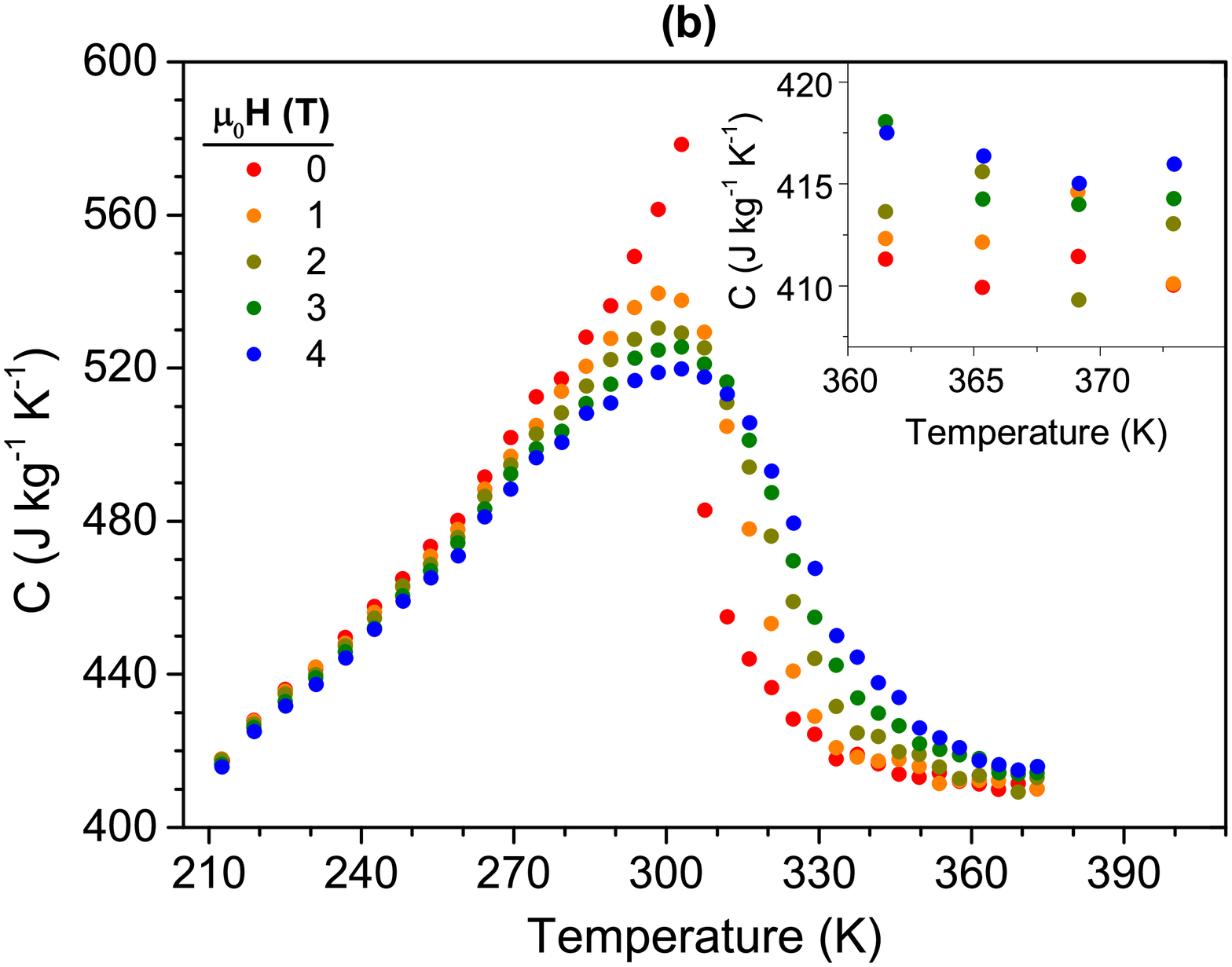,width=8cm,clip=}
\caption{(a) Temperature dependence of the specific heat of Ni$_{50}$Mn$_{35.5}$In$_{14.5}$ measured  in the absence of magnetic field and under an applied field of 2 T (as indicated by the colour code in the figure). Continuous and discontinuous lines correspond to heating and cooling runs, respectively. (b) Temperature dependence of the specific heat of Ni$_{50}$Mn$_{34}$In$_{16}$ measured using a relaxation calorimeter (PPMS from Quantum Design) under magnetic fields of 0, 1, 2, 3 and 4 T (as indicated by the colour code in the figure.)}
\label{fig4}
\end{figure*}

Using  data for $C^M$ and $C^A$, and the base-line corrected calorimetric curves (Fig. 3), we have computed the entropy curves $S(T, \sigma , H)$ as follows: 

\begin{widetext}
\begin{equation}
S(T,\sigma , H) = \left\lbrace
\begin{array}{l} \vspace{0.2cm}
\displaystyle\int_{T_0}^{T} \frac{C^{M} (T, \sigma, H)}{T} dT \hspace{1cm} ;  \hspace{2cm}  T \leq T_1 \\ \vspace{0.2cm}
S(T_1,\sigma, H) + \displaystyle\int_{T_1}^{T} \frac{1}{T} \left(C (T, \sigma, H) + \frac{dQ(T,\sigma , H)}{dT} \right) dT \hspace{0.3cm} ; \hspace{0.2cm}  T_1 < T \leq T_2 \\ \vspace{0.2cm}
S(T_2,\sigma, H) + \displaystyle\int_{T_2}^{T}  \frac{C^{A} (T, \sigma, H)}{T} dT \hspace{1cm} ; \hspace{0.5cm}  T_2 < T \\ 
\end{array}
\right.
\end{equation}
\end{widetext}

\noindent where $C = x C^M + (1-x) C^A$, with $x$ the fraction of the sample which has transformed to martensite. Results for selected values of magnetic field and uniaxial stress are shown in Figs. 5(a-c) and 6(a-c).

\begin{figure*}
\epsfig{file=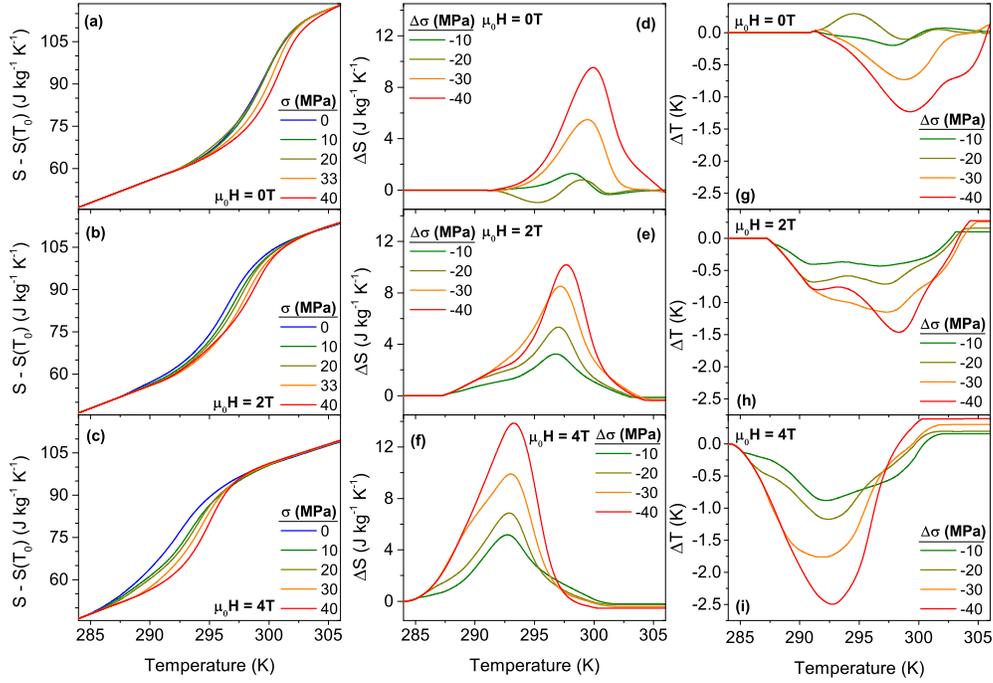,width=14cm}
\caption{(a)-(c) Entropy curves (referenced to a value at $T_0$ = 256 K in the absence of uniaxial stress and magnetic field) at constant values of uniaxial stress and magnetic field, (d)-(f) elastocaloric isothermal entropy changes and (g)-(i) elastocaloric adiabatic temperature changes corresponding to the removal of a uniaxial stress. Each panel corresponds to a constant applied magnetic field (indicated by the label in the panel), and the value of the uniaxial stress is indicated by the colour code.}
\label{fig5}
\end{figure*}

\begin{figure*}
\epsfig{file=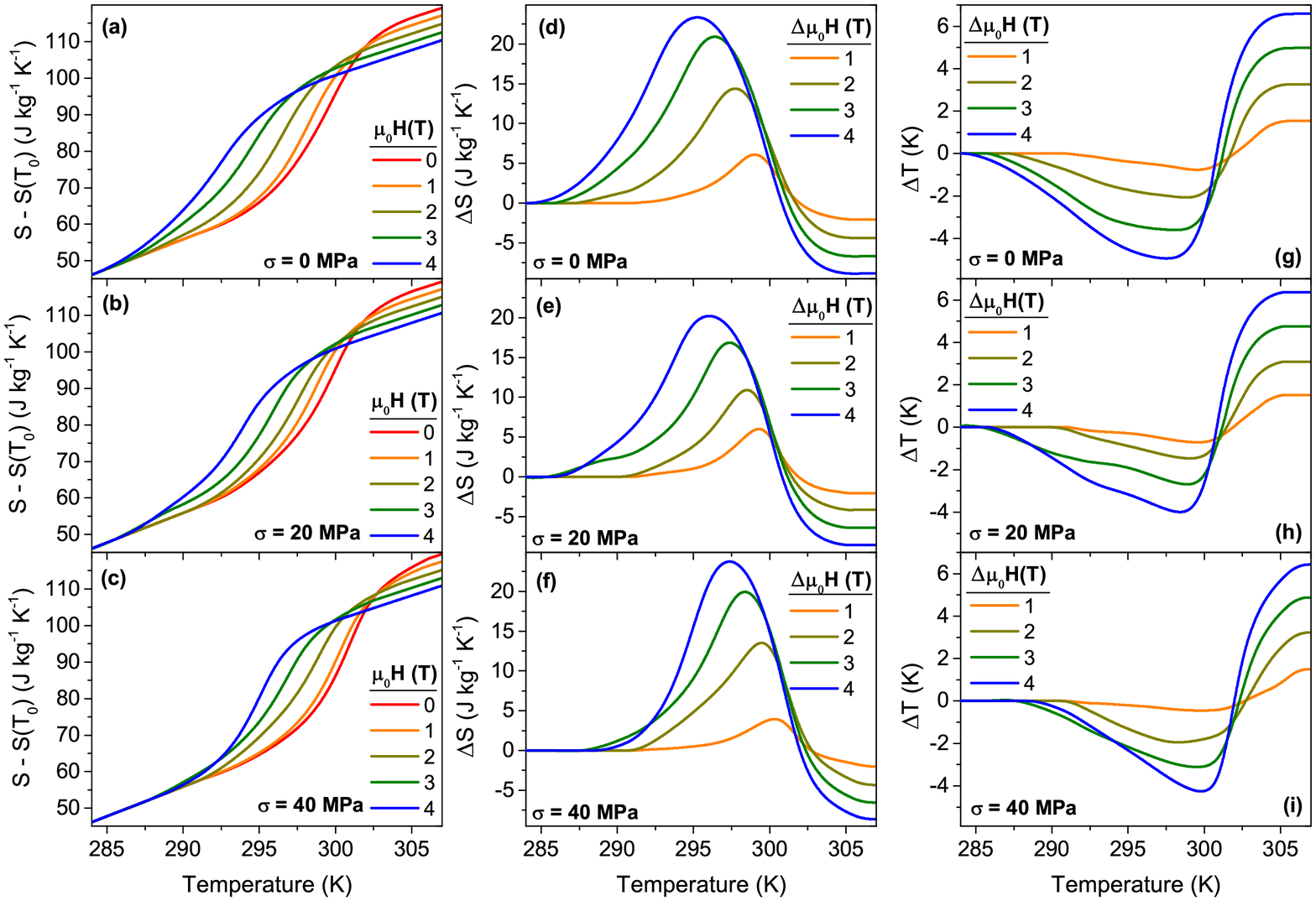,width=14cm}
\caption{(a)-(c) Entropy curves (referenced to a value at $T_0$ = 256 K in the absence of uniaxial stress and magnetic field) at constant values of uniaxial stress and magnetic field, (d)-(f) magnetocaloric isothermal entropy changes and (g)-(i) magnetocaloric adiabatic temperature changes corresponding to the application of a magnetic field. Each panel corresponds to a constant applied uniaxial stress (indicated by the label in the panel), and the value of the magnetic field is indicated by the colour code.}
\label{fig6}
\end{figure*}

All $S(T, \sigma, H)$ values are referenced to the entropy at $T_{0}$ = 256 K, in the absence of stress and magnetic field. For each specific value of magnetic field, the entropy curves shift towards higher temperatures with increasing stress, and they join in the high-temperature region where entropy values coincide thereby indicating a stress-independent entropy in the austenite (Figs. 5(a-c)). On the other hand, for any fixed value of applied stress, the entropy curves shift to lower temperatures with increasing magnetic field, and they exhibit a cross-over: in the austenitic phase the entropy  decreases with increasing magnetic field (Figs. 6(a-c)). Such a decrease reflects the   decrease in $\Delta S_t$ (see Fig. 3(f)) and, as previously mentioned, is a consequence of the increase  in the magnetic contribution with increasing magnetic field.

\subsection{Elastocaloric and magnetocaloric effects. Isothermal entropy and adiabatic temperature changes}

We have computed the isothermal entropy changes ($\Delta S$) and adiabatic temperature changes ($\Delta T$) corresponding to the elastocaloric and magnetocaloric effects in Ni$_{50}$Mn$_{35.5}$In$_{14.5}$. Because we have used calorimetric curves for heating runs, the studied caloric effects correspond to the transition from martensite to austenite. It is to be noted that the application of magnetic field stabilizes austenite, whereas the application of uniaxial stress stabilizes martensite. Furthermore, due to the hysteresis of the transition, the application of both a magnetic field and mechanical stress will take the sample through a minor loop within the two-phase coexistence region (for which no experimental data are available). For these reasons, the analysis of single caloric and multicaloric effects will be restricted to values obtained from the application of the magnetic field and the removal of stress (they correspond to trajectories from below  to above the red plane in Fig. 2(a)).

\begin{figure*}
\epsfig{file=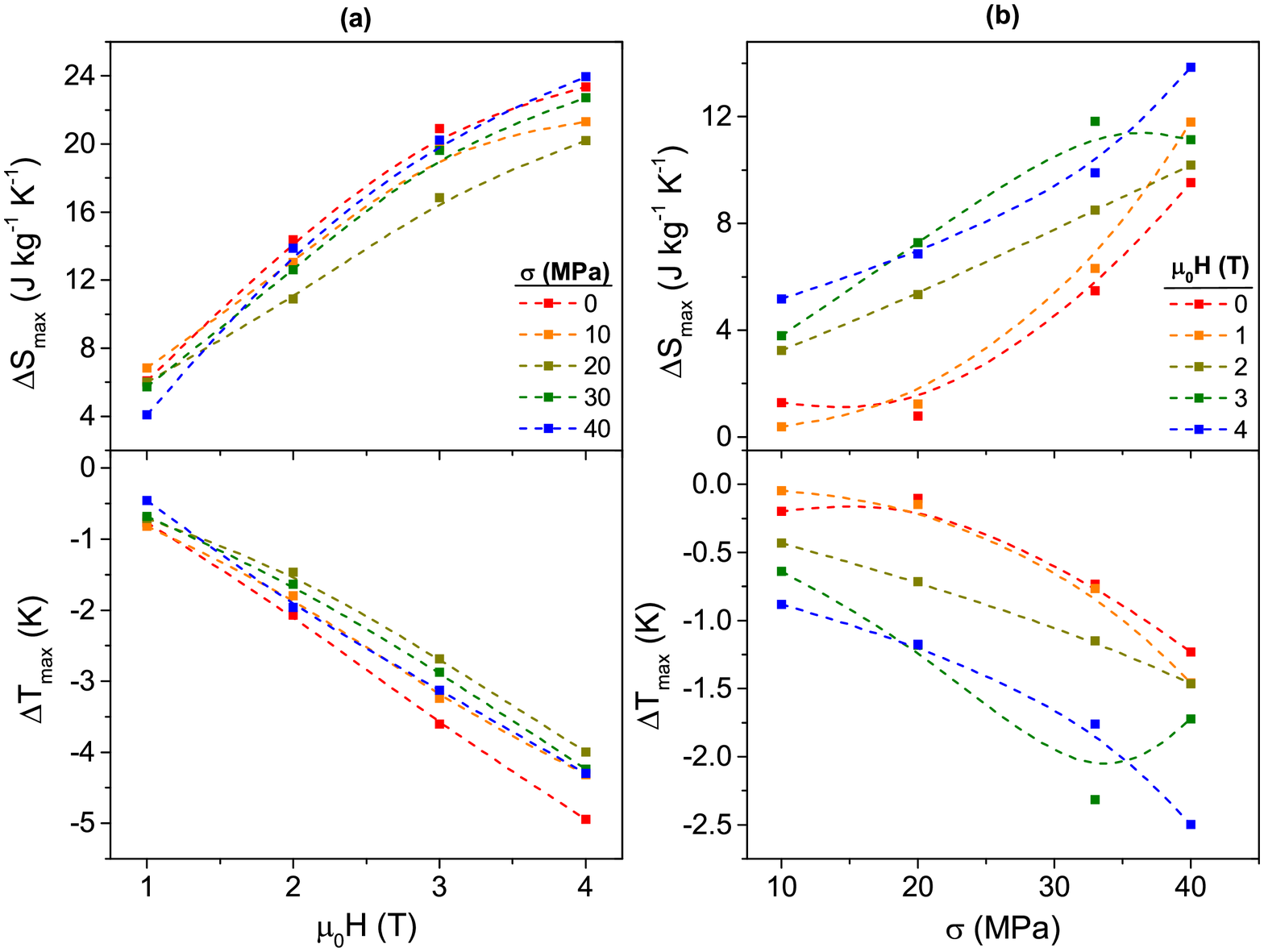,width=14cm,clip=}
\caption{(a) Magnetic-field dependence of the extreme values for the magnetocaloric isothermal entropy change and adiabatic temperature change  for different values of applied (constant) uniaxial stress. (b) Stress dependence of the extreme values for the elastocaloric isothermal entropy change  and adiabatic temperature change  for different values of applied (constant) magnetic field. Lines are guides to the eye.}
\label{fig7}
\end{figure*}

From $S(T,\sigma , H)$, the isothermal entropy change ($\Delta S $) associated with a given caloric effect arising from the application of a magnetic field $H$, and the stress removal from a value $\sigma$ is obtained as

\begin{equation}
\Delta S(T,\sigma \rightarrow 0, 0  \rightarrow H) = S(T,0,H) - S(T,\sigma, 0) 
\end{equation}

The adiabatic temperature change ($\Delta T$) is computed from the  $T(S,\sigma , H)$ (obtained by inversion of the corresponding entropy curves) as

\begin{equation}
\Delta T(S,\sigma \rightarrow 0, 0  \rightarrow H) = T(S,0,H) - T(S,\sigma, 0).
\end{equation}

\noindent The previous expression provides $\Delta T$ as a function of entropy, but it is customary to plot adiabatic temperature changes as a function of temperature. Such a temperature dependence is easily obtained by plotting each $\Delta T$ data at the temperature given by the $S(T,\sigma , H)$ curve, i.e. the temperature  prior to the application (removal) of magnetic field (stress).   

Results for  $\Delta S (T, \sigma \rightarrow 0, H)$
and $\Delta T (T, \sigma \rightarrow 0, H)$, $\Delta S (T, \sigma,  0 \rightarrow H)$ and $\Delta T (T, \sigma , 0 \rightarrow H)$, are shown as a function of temperature in  Figs. 5(d-f), 5(g-i), 6 (d-f) and 6 (g-i), respectively. For the elastocaloric effect (in the absence of magnetic field), the shift of the transition is very small at low stresses and the observed small fluctuations of $\Delta S$ and $\Delta T$ curves around zero  cannot be considered to be physically meaningful within experimental errors. The elastocaloric effect has been found to be conventional while a cross-over from inverse to conventional is observed for the magnetocaloric effect. The inverse magnetocaloric effect at lower temperatures arises from the martensitic transition while the conventional magnetocaloric effect at higher temperatures is associated with changes in the ferromagnetic order in the vicinity of the Curie point of the austenite. In all cases, an increase in the magnitude of the external stimulus (magnetic field for the magnetocaloric effect and stress for the elastocaloric effect) enlarges the temperature window of the corresponding caloric effect.

Fig. 7 shows the extreme values for the isothermal entropy change ($\Delta S_{max}$) and for the adiabatic temperature change ($\Delta T_{max}$) as a function of magnetic field (at constant applied stress) for the magnetocaloric effect (Fig. 7(a)) and as a function of released stress (at constant applied magnetic field) for the elastocaloric effect (Fig. 7(b)). For any applied load, the magnetocaloric  $\Delta S_{max}$ and $|\Delta T_{max}|$  increase for higher fields. The increase in $|\Delta T_{max}|$ is found to be linear but $\Delta S_{max}$ values show a tendency to saturate at high fields. The fact that $\Delta S_{max}$ values are lower than the transition entropy change ($\Delta S_t (H)$,  Fig. 3(f)) indicates that for the studied range of magnetic fields the sample does not completely transform. For the elastocaloric effect, $\Delta S_{max}$ and $|\Delta T_{max}|$ values are more scattered but both quantities increase with increasing $|\sigma|$. Again, $\Delta S_{max}$ values are lower than $\Delta S_t$ showing that larger stresses are required to fully transform the sample. In relation to the effect of a secondary field, no systematic dependence for the magnetocaloric $\Delta S_{max}$ and $|\Delta T_{max}|$ with stress has been found within the range of studied stresses. In spite of the scatter of the data, elastocaloric $\Delta S_{max}$ and $|\Delta T_{max}|$ increase for higher applied magnetic fields. Although this result seems to be in contrast with the decrease of the transition entropy change ($\Delta S_t$) with increasing magnetic field, it must be taken into account that magnetic field shifts the martensitic transition to lower temperatures. Therefore, under an applied magnetic field the fraction of the sample in the martensitic state is larger than in the absence of the field (for  given values of temperature and applied stress). Hence, when the stress is released, the larger transformed fraction gives rise to a larger elastocaloric $\Delta S_{max}$.

While our method based on DSC under external fields provides reliably accurate data for the entropy and temperature changes for caloric effects arising from first-order phase transitions, it is less suited to study caloric effects around continuous phase transitions. For that reason, although our data correctly captures the cross-over of the magnetocaloric effect from inverse to conventional, actual data for $\Delta S$ and $\Delta T$ associated with the conventional magnetocaloric effect around $T_c$ might be inaccurate. With the aim of corroborating the cross-over from the inverse to the conventional magnetocaloric effect, and to confirm the temperature region where the inverse magnetocaloric effect takes place, we have performed direct measurements of the adiabatic temperature change resulting from the application of a 1.64 T magnetic field in the absence of stress and for an applied (constant) uniaxial stress of 40 MPa. Results are shown in the Supplementary Material  (Fig. S1) in comparison with data derived from our entropy curves (which will be described later in this paper). The comparison between direct and indirect data provides indeed a confirmation of the temperature region where the inverse  magnetocaloric effect occurs (including the cross-over temperature from inverse to conventional magnetocaloric effect), and also of the shift in the magnetocaloric effect with applied uniaxial stress.   

\subsection{Entropy surfaces.}

Experimental entropy curves (Figs. 5 (a-c) and 6 (a-c)) are only known for given values of magnetic field and stress. In order to obtain the multicaloric effect over the entire ($T$,$\sigma$,$H$) thermodynamic phase space, it is necessary to fit a numerical function $S(T,\sigma,H$) to the experimental data. 

We have used the following function to fit the entropy curves

\begin{widetext}
\begin{equation}
S(T,\sigma , H) - S(T_{0}) = \int_{T_{0}}^T \frac{C(T, \sigma, H)}{T} dT + \Delta S_{t}(\sigma , H) X(T,\sigma , H)
\end{equation}
\end{widetext}

\noindent with $T_0$ = 256 K. 

The fitting function $X(T,\sigma , H)$ is taken as:

\begin{equation}
X(T,\sigma , H) = 1 - \frac{1}{e^{B(T-T_t)} + 1}
\end{equation}

\noindent where  $B$ and $T_t$ are free parameters of the fit which depend on $H$ and $\sigma$, and the best fit to experimental data is found for


%



%
%

\begin{eqnarray}
\nonumber 
B(H,\sigma) = 0.74(2) - 0.032(18) H + 0.0019(17) \sigma \\ 
\nonumber 
- 0.004(4) H^2  - 2 \times 10^{-5} (4) 
\sigma^2 - 0.2 \times 10^{-4} (4) H \sigma 
\end{eqnarray}

\noindent and

\begin{eqnarray}
\nonumber
T_{t}(H,\sigma) = 299.1(2) - 1.14(14) H + 0.016(15) \sigma \\ 
\nonumber
- 0.15(3) H^2  + 4 \times 10^{-4} (3)
 \sigma^2 + 0.008(3) H \sigma
\end{eqnarray}

Fig. 8 illustrates the agreement between the fitted (dashed lines) and experimental data (solid lines) for selected values of stress and magnetic field.

\begin{figure*}
\epsfig{file=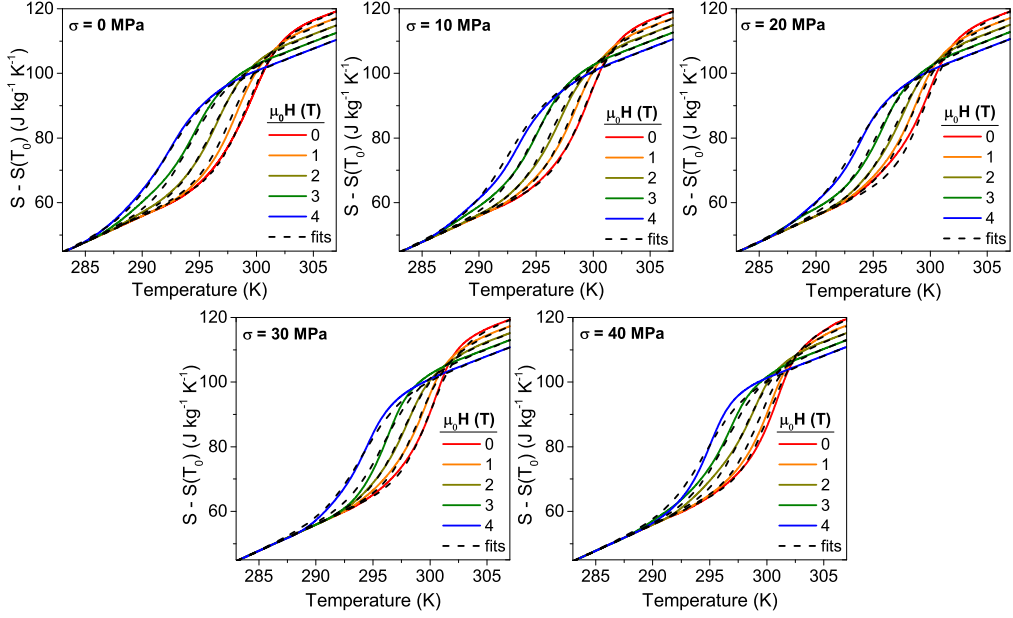,width=14cm,clip=}
\caption{Temperature dependence of the entropy (referenced to its value at $T_0$ = 256 K in the absence of stress and magnetic field) for selected values of applied (constant) uniaxial stress (indicated by the labels in each panel) and magnetic field (indicated by the colour code). Solid lines correspond to the experimental data, and dashed lines, to the fitted curves.}
\label{fig8}
\end{figure*}

From the fitted entropy curves $S(T,\sigma,H)$, the elastocaloric effect (under an applied magnetic field), $\Delta S (T, \sigma \rightarrow 0, H)$, the  magnetocaloric effect (under applied uniaxial stress) $\Delta S (T, \sigma , 0 \rightarrow H)$, and multicaloric effects $\Delta S (T, \sigma \rightarrow 0, 0 \rightarrow H)$ are computed as:

\begin{equation}
\Delta S(T, \sigma \rightarrow 0, H) = S(T, 0, H) - S(T, \sigma, H)   
\end{equation}

\begin{equation}
\Delta S(T, \sigma, 0 \rightarrow H) = S(T, \sigma, H) - S(T, \sigma, 0)   
\end{equation}

\begin{equation}
\Delta S(T, \sigma \rightarrow 0, 0 \rightarrow H) = S(T, 0, H) - S(T, \sigma, 0)
\end{equation}

\noindent where the above expressions have to be computed numerically. 

\begin{figure*}
\epsfig{file=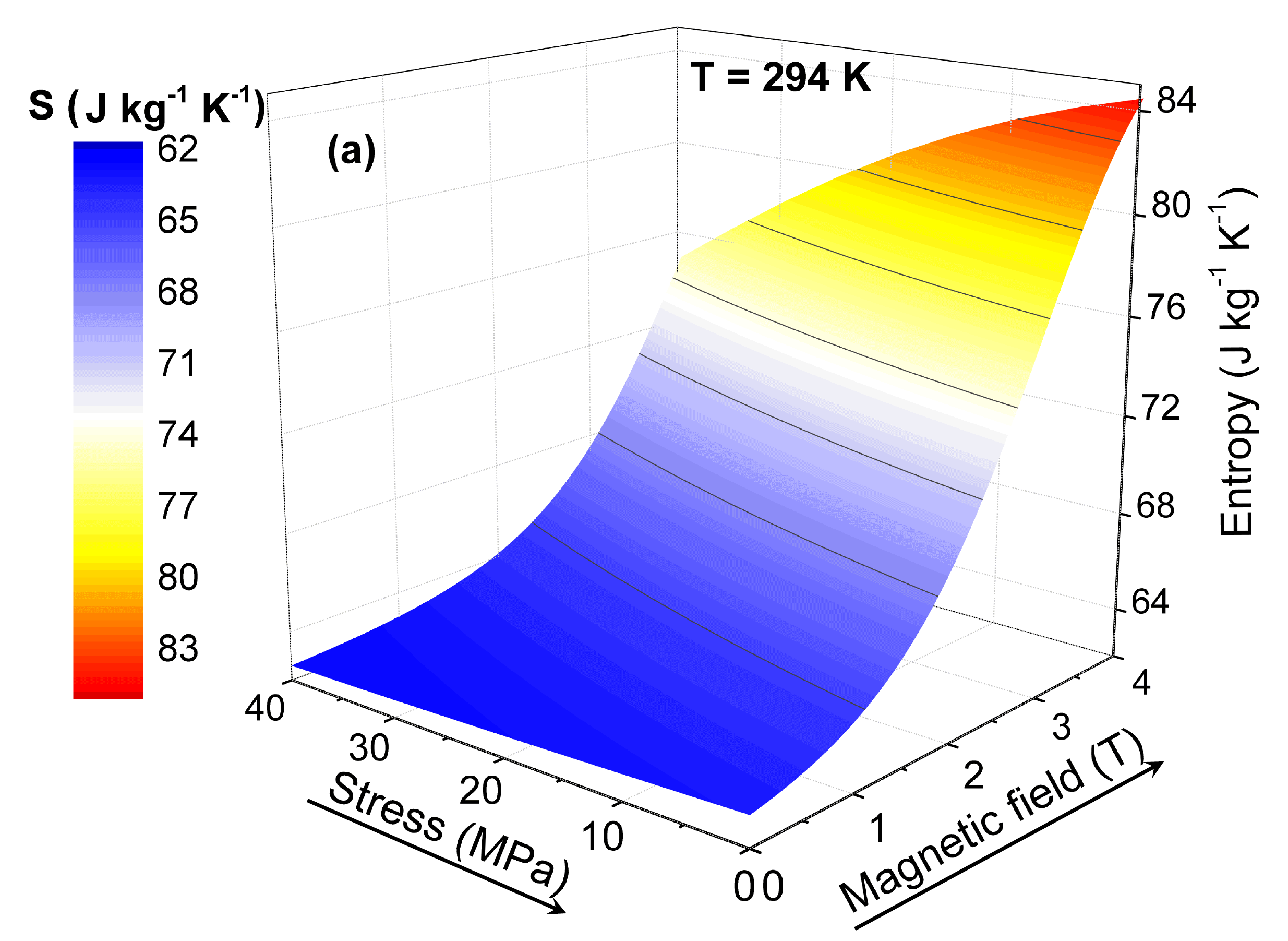,width=7cm,clip=}
\epsfig{file=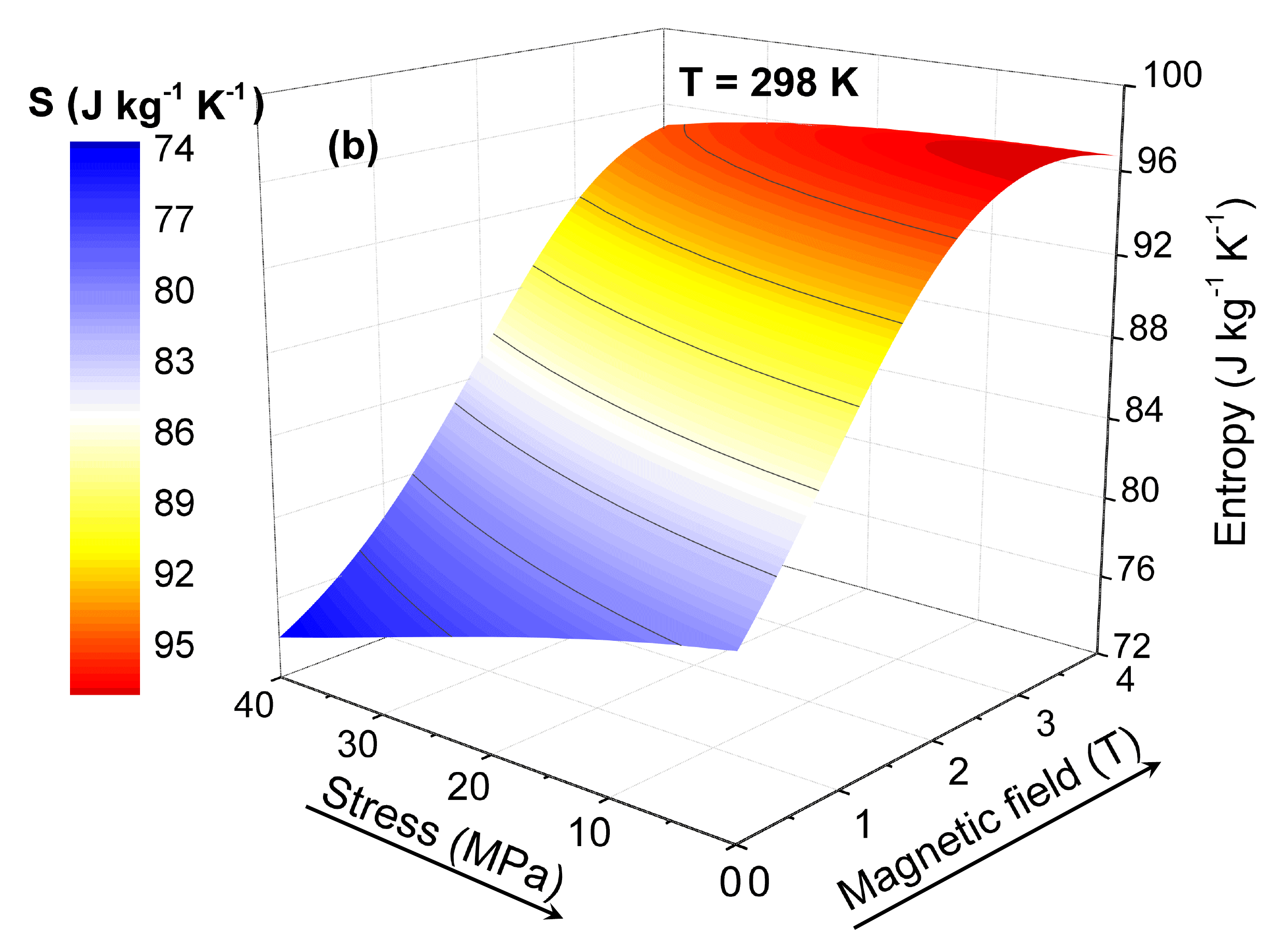,width=7cm,clip=}
\epsfig{file=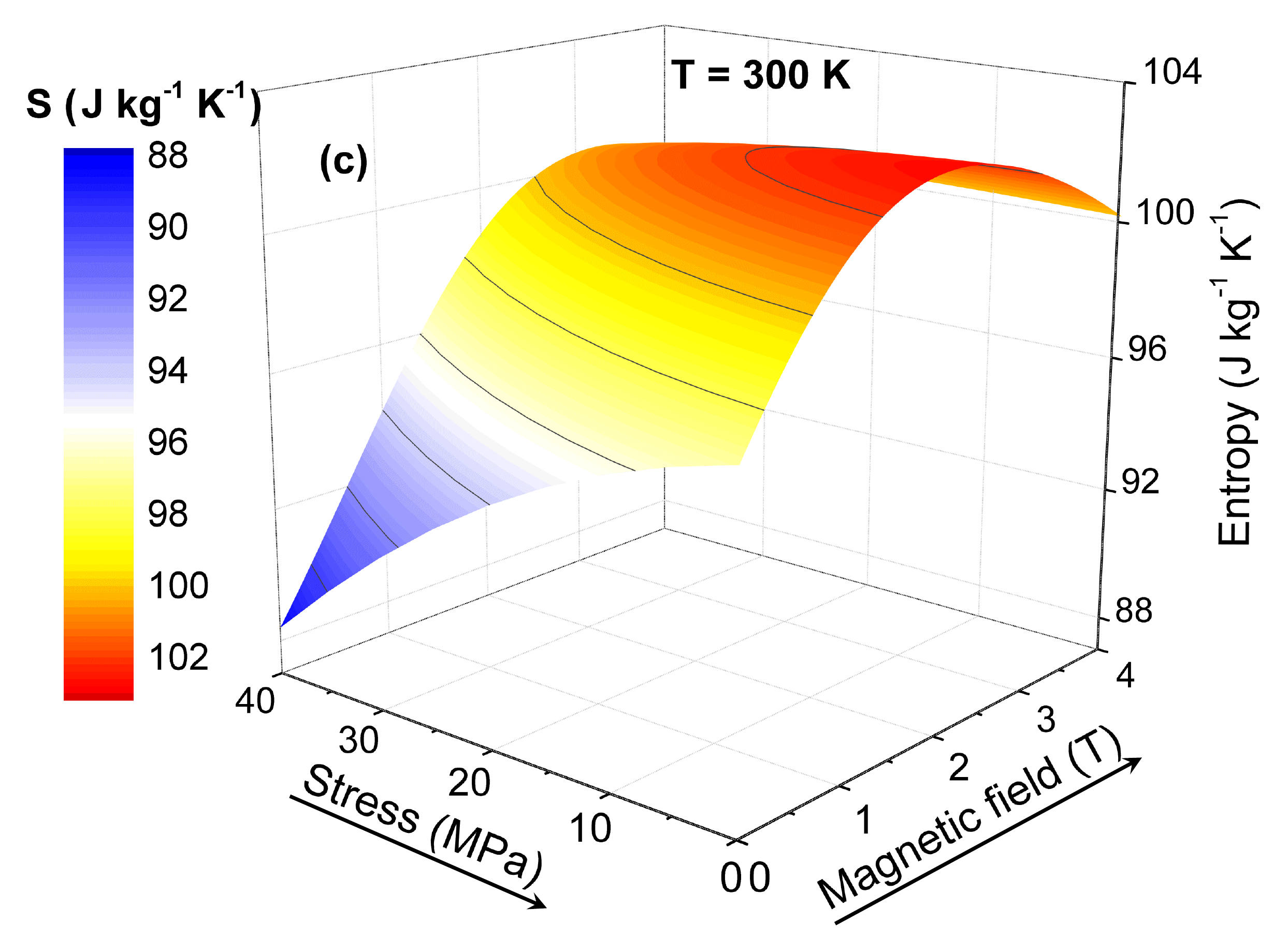,width=7cm,clip=}
\epsfig{file=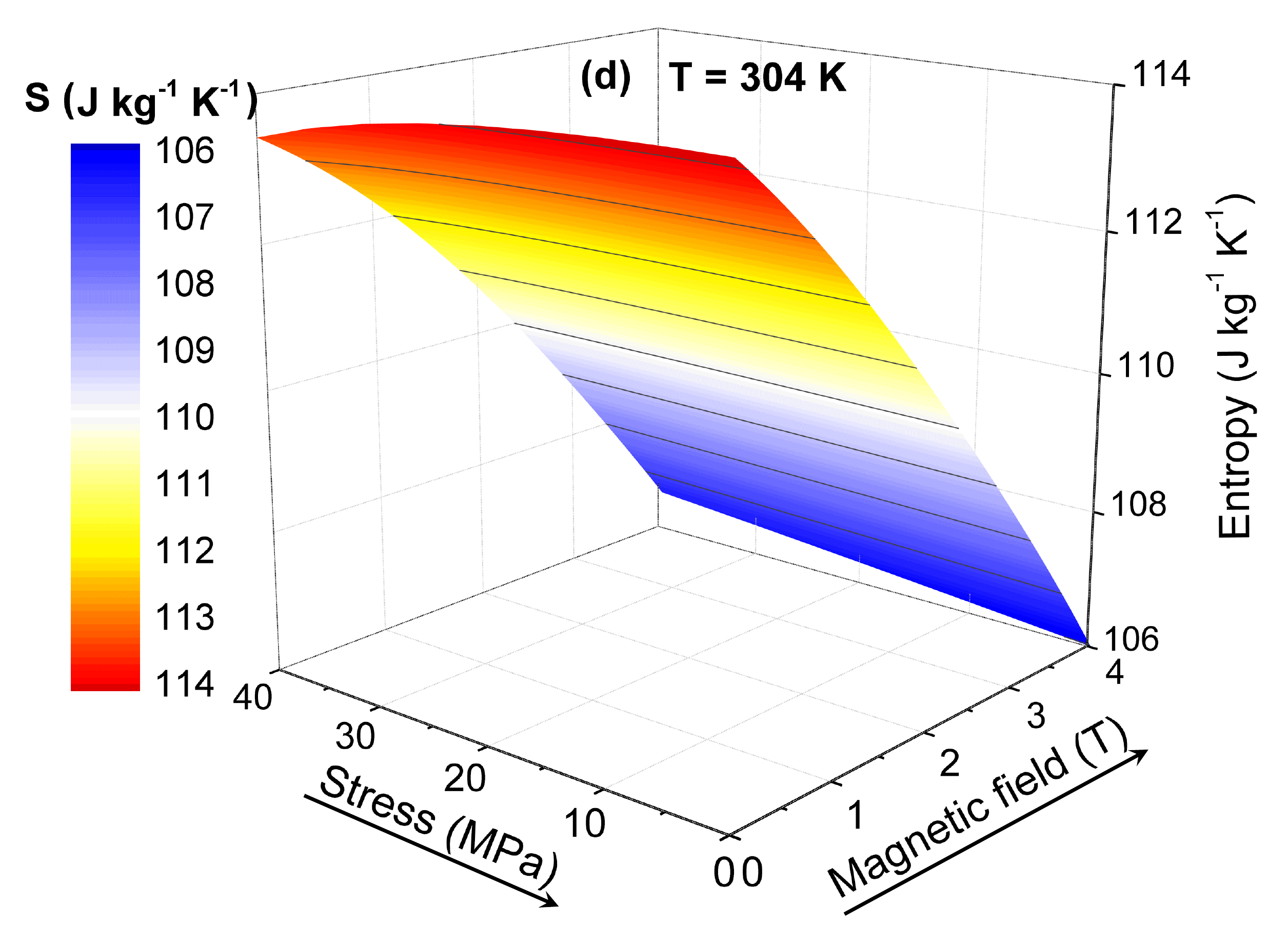,width=7cm,clip=}
\epsfig{file=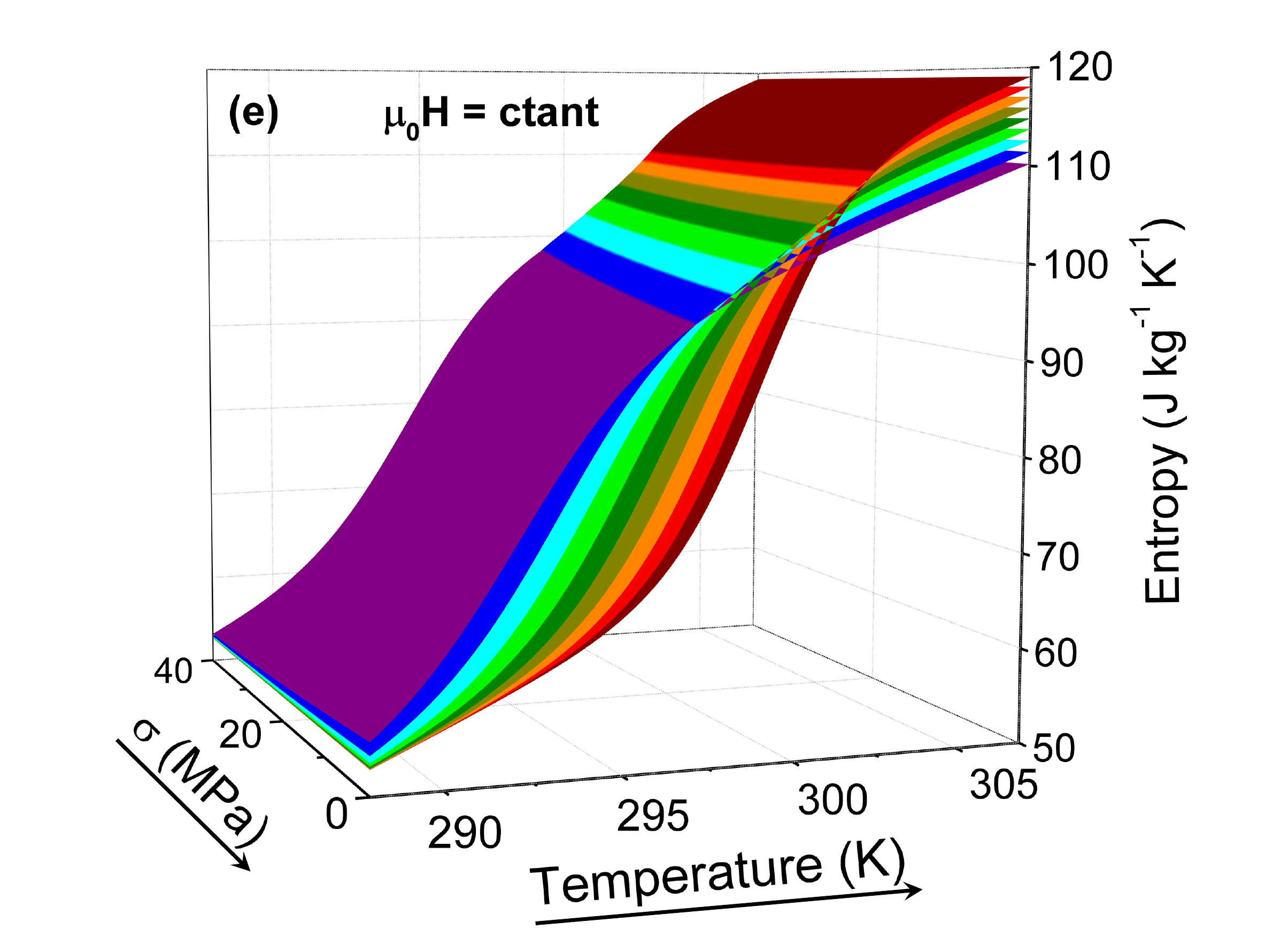,width=7cm,clip=}
\epsfig{file=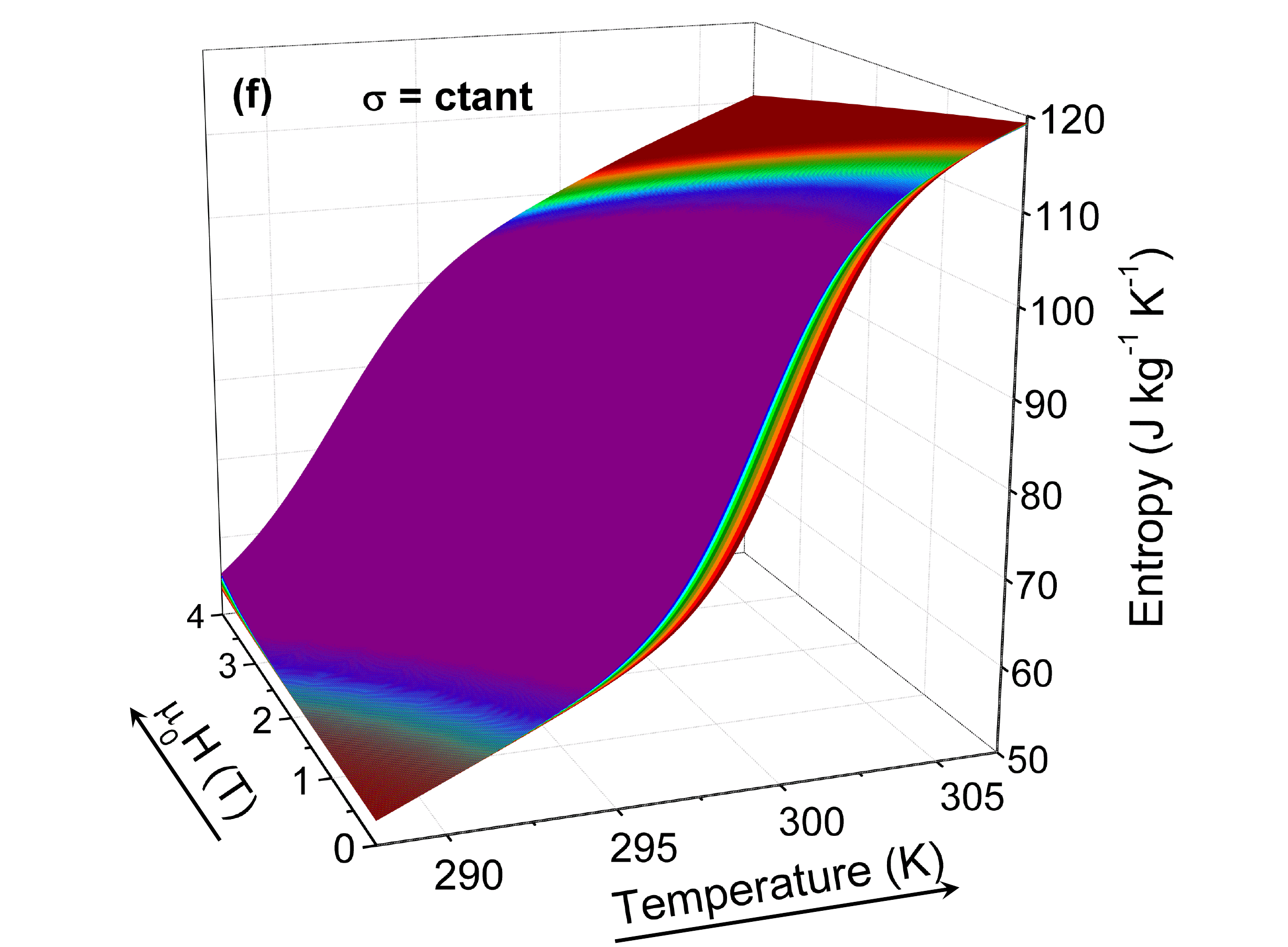,width=7cm,clip=}
\caption{(a)-(d) Isothermal entropy surfaces as a function of magnetic field and stress. (e) Isofield entropy surfaces as a function of stress and temperature. Each surface corresponds to a fixed magnetic field from 0 to 4 T (steps of 0.5 T). (f) Isostress entropy surfaces as a function of temperature and magnetic field. Each surface corresponds to a fixed uniaxial stress from 0 to 40 MPa (steps of 5 MPa). In all cases, the entropy is referenced to a value at $T_0$= 256 K in the absence of magnetic field and stress.}
\label{fig9}
\end{figure*}

Isothermal entropy surfaces are plotted as a function of stress and magnetic field, at selected values of temperature in Figs. 9(a-d). For temperatures at the onset of the reverse martensitic transition, $S(T,\sigma,H)$ increases with increasing magnetic field, and there is also a tiny increase with decreasing stress. For temperatures within the transition region, the $S(T,\sigma,H)$ increase with decreasing stress is more pronounced. On the other hand, $S(T,\sigma,H)$  increases with increasing magnetic field up to a maximum, and decreases for larger  fields. The value of the magnetic field where this maximum occurs is temperature dependent. At temperatures above the reverse martensitic transition, the entropy decreases with increasing magnetic field, in accordance with the ferromagnetic nature of the austenitic phase, while no significant stress dependence is observed. Figures 9(e) and 9(f) show isofield and isostress entropy surfaces plotted as a function of stress and temperature (Fig. 9(e)), and as a function of magnetic field and temperature (Fig. 9 (f)). It is worthwhile to remember that due to the hysteresis of the transition, these surfaces are only representative for increasing field and decreasing stress (as indicated by arrows in the axis of the figures). The cross-over behaviour for the magnetocaloric effect is evident within all the range of temperatures and stresses under study (Fig. 9(e)). Furthermore, the marked decrease in entropy as magnetic field increases together with the weak dependence in stress is also apparent from Fig. 9(f). 

\begin{figure*}
\epsfig{file=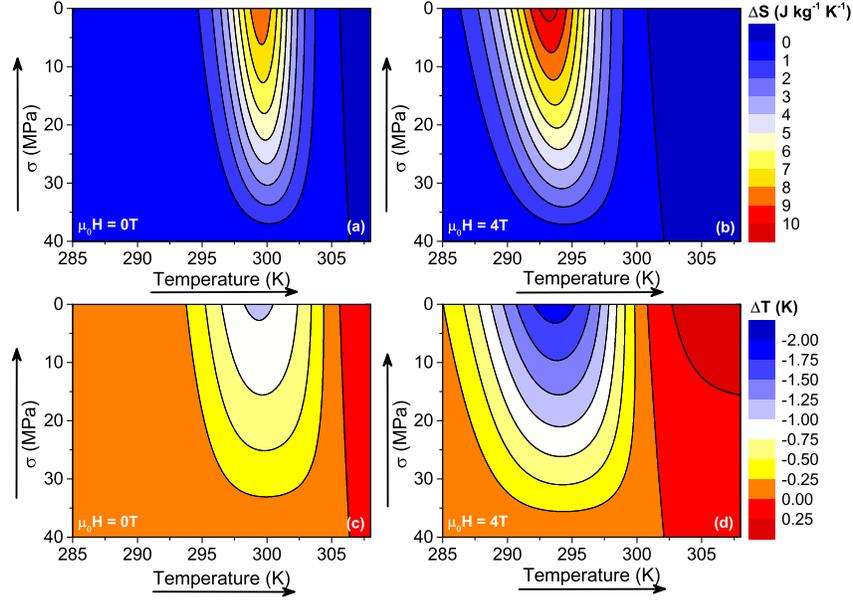,width=12cm,clip=}
\caption{Contour colour plots corresponding to the elastocaloric isothermal entropy changes (a and b) and adiabatic temperature changes (c and d) under the removal of a uniaxial stress from an initial value of 40 MPa (as indicated by the arrows in the stress axis). Left panels (a and c) correspond to values without applied magnetic field, and right panels (b and d) to data under an applied constant magnetic field of 4 T.}
\label{fig10}
\end{figure*}

\begin{figure*}
\epsfig{file=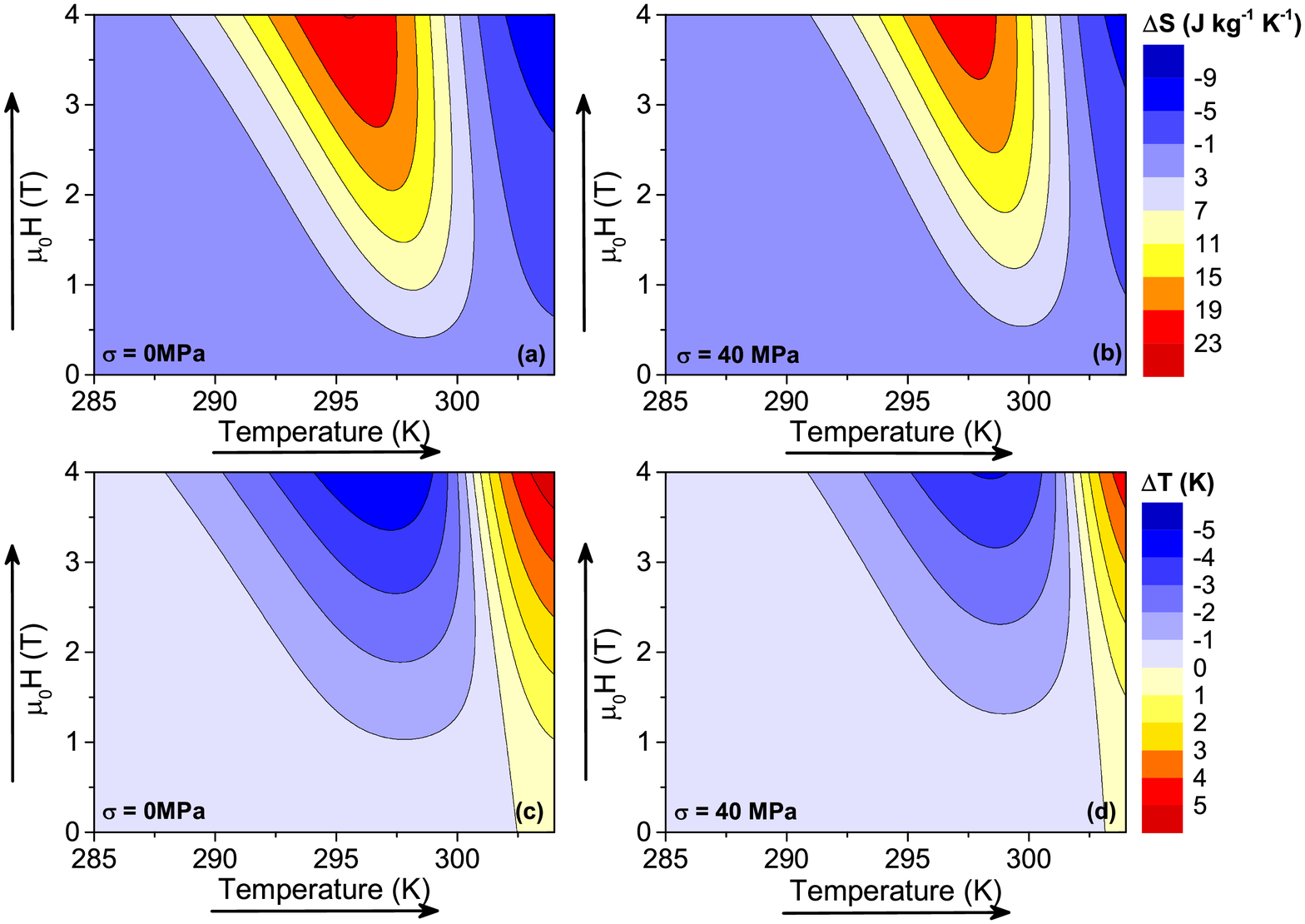,width=12cm,clip=}
\caption{Contour colour plots corresponding to the magnetocaloric isothermal entropy changes (a) and (b) and adiabatic temperature changes (c) and (d) under the application of a magnetic field (as indicated by the arrows in the field axis). Left panels (a) and (c) correspond to values without applied uniaxial stress, and right panels (b) and (d) to data under an applied constant uniaxial stress of 40 MPa.}
\label{fig11}
\end{figure*}

From our fitted $S(T,\sigma ,H)$ function it is possible to separately compute single caloric and multicaloric effects. Single caloric effects while a constant secondary field is applied are shown in  Figs. 10 and 11 as colour contour plot maps, for the   elastocaloric effect (under an applied $H$) (Fig. 10)   and for the magnetocaloric effect (under an applied $\sigma$) (Fig. 11).  The corresponding three-dimensional plots are shown in Figs. S2 and S3 of the Supplementary Material. Overall, there is good agreement between the isothermal entropy and adiabatic temperature changes obtained from the fitted $S(T,\sigma, H)$ functions with the corresponding data computed from the raw entropy curves (Figs. 5 and 6). Such an agreement confirms the robustness of our fitting procedure and provides confidence to the multicaloric data that will be later derived from the fitted $S(T,\sigma, H)$ functions.

With regards to the elastocaloric effect (Fig. 10), the application of a magnetic field enlarges  the temperature window where the giant elastocaloric effect occurs and shifs this window to lower temperatures. Furthermore, the application of a magnetic field enchances the elastocaloric effect: the elastocaloric $\Delta S$ and $\Delta T$ obtained under an applied magnetic field are larger than those in the absence of field. In particular, a stress removal from 40 MPa to zero, under a magnetic field of 4 T renders $\Delta S$ = 10.4 J kg$^{-1}$ K$^{-1}$ and $\Delta T$ = - 1.9 K, while in the absence of magnetic field $\Delta S$ = 8.7 J kg$^{-1}$ K$^{-1}$ and $\Delta T$ = - 1.0 K.

In relation to the magnetocaloric effect (Fig. 11), the influence of uniaxial stress is very weak (due to the low range of applied stresses). There is a slight shift of the region where the giant (inverse) magnetocaloric effect occurs towards higher temperatures under an applied uniaxial stress. Furthermore, the overall magnetocaloric effect slightly shifts to higher magnetic fields when stress is applied. For instance, if we focus on the contour plot corresponding to $\Delta S \sim$  19 J kg$^{-1}$ K$^{-1}$,  it is observed that a magnetic field of 2.7 T is needed in the stress free case (Fig. 11(a)), while a magnetic field of 3.3 T is necessary to achieve this value when a 40 MPa stress is applied (Fig. 11(b)). This shift towards higher magnetic field values is due to the fact that uniaxial stress stabilizes the martensitic phase. The maximum magnetic-field induced entropy change is $\Delta S$ = 23.1 kg$^{-1}$ K$^{-1}$, corresponding to a magnetic field of 4 T and the absence of applied stress.

\subsection{Multicaloric effects. Isothermal entropy and adiabatic temperature changes}

\begin{figure*}
\epsfig{file=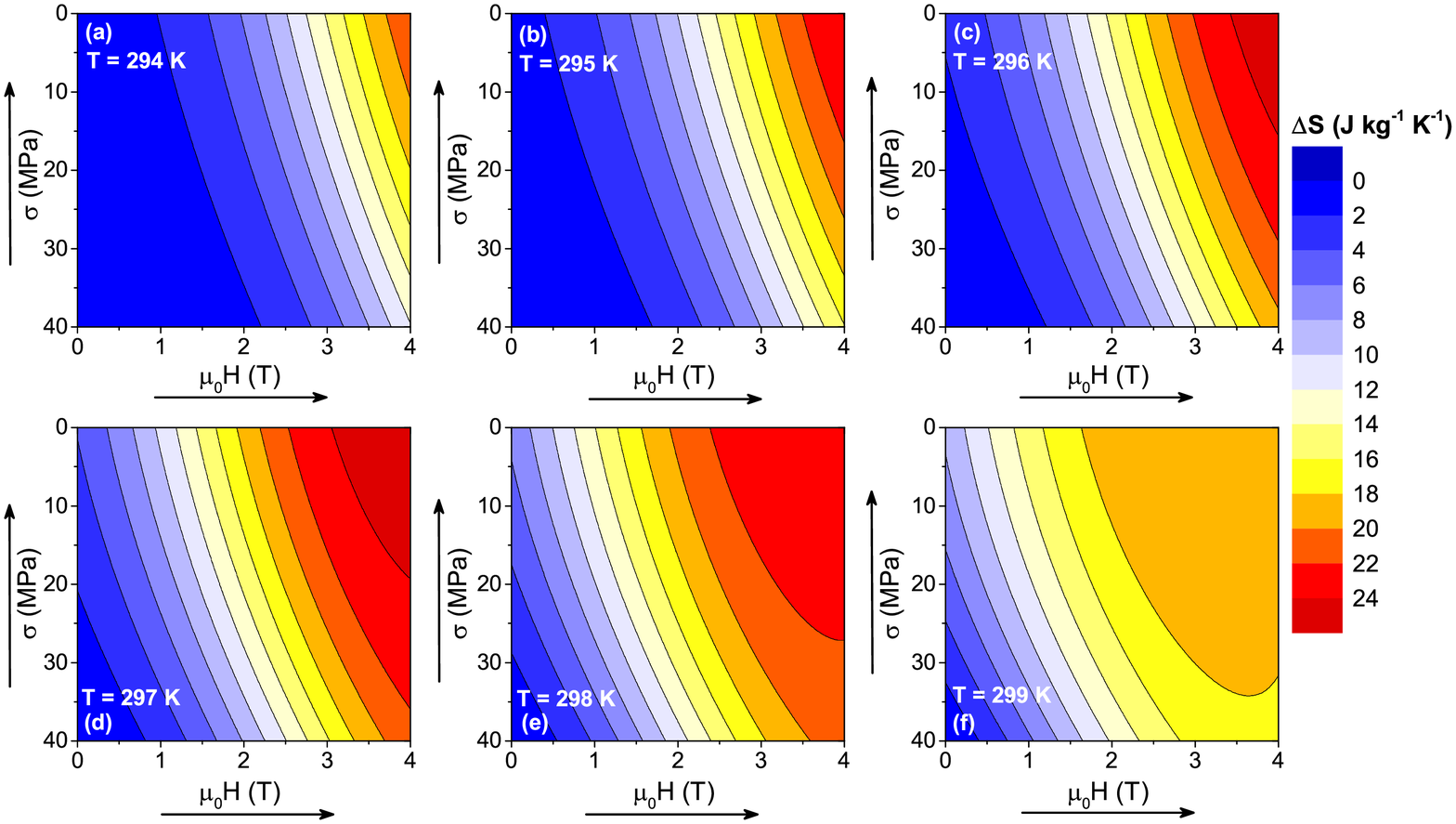,width=14cm}
\caption{Colour contour plots for the multicaloric isothermal entropy change resulting from the application of a magnetic field and the removal of a uniaxial stress (from an initial stress of 40 MPa).}
\label{fig12}
\end{figure*}

\begin{figure*}
\epsfig{file=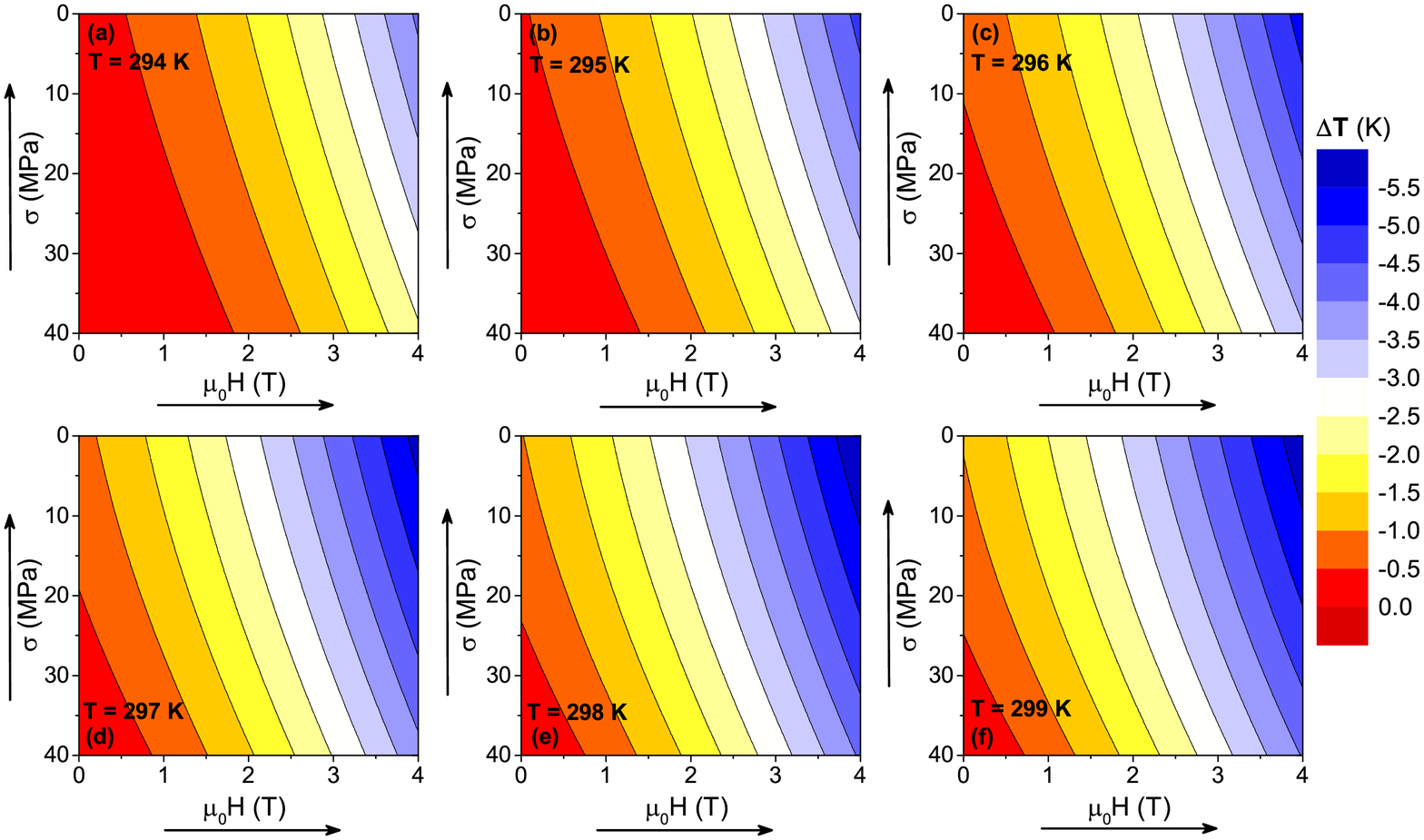,width=14cm}
\caption{Colour contour plots for the multicaloric adiabatic temperature change resulting from the application of a magnetic field and the removal of a uniaxial stress (from an initial stress of 40 MPa).}
\label{fig13}
\end{figure*}

The multicaloric effect refers to the isothermal entropy and adiabatic temperature changes when two (or more) external fields are modified. Because both quantities are state function, no distinction is made on whether they are applied (or removed) simultaneously or sequentially \cite{SternTaulats2018}.
The multicaloric response of our system when both uniaxial stress and magnetic field are varied can be computed from the isothermal entropy surfaces in the $H$  and $\sigma$ parameter space  for selected values of temperature (Figs. 9(a)-(d)).  The multicaloric isothermal entropy change for an arbitrary change in $H$ and $\sigma$ is simply obtained by subtracting the value at the origin $S(T,\sigma = $40  MPa$ ,0)$ to the actual value on the surface $S(T,\sigma, H)$. The multicaloric adiabatic temperature change can be computed in a similar way from the inverted entropy functions $T(S, \sigma, H)$. Results for the multicaloric $\Delta S$ and $\Delta T$ are shown  in Figs. 12 and 13 as colour contour maps. They show a clear improvement in the caloric response of the studied alloy by the combined effect of the two external stimuli. On the one hand, the maximum values achieved for the multicaloric entropy change ($\Delta S = $ 25.2 J kg$^{-1}$ K$^{-1}$ at $T$ = 296 K and $\Delta S = $ 24.9 J kg$^{-1}$ K$^{-1}$ at $T$ = 297 K ) clearly exceed the values for single magnetocaloric and elastocaloric effects.  Furthermore, such  maximum values can be obtained within a certain range of stress and magnetic field. In particular, at 297 K (Fig. 12(d)) entropy values equal or larger than 24 J kg$^{-1}$ K$^{-1}$ are obtained for the combination of $H$ and $\sigma$ ranging in a window limited by the following values: $\mu_ 0 H$ (0 $\rightarrow$ 4T), $\sigma$ (40 MPa $\rightarrow$ 20 MPa), and $\mu_0 H$ (0 $\rightarrow$ 3.1 T), $\sigma$ (40 MPa $\rightarrow$ 0). On the other hand, it is possible to achieve large entropy changes at relatively low values of the applied external field. For instance, if we focus on a magnetic field of 1 T (which is readily accessible by permanent magnets), it is found that the entropy values corresponding to the single caloric effect are in the range 4-6 J kg$^{-1}$ K$^{-1}$ (see Fig. 7(a)). However, these values can be doubled by combining the application of 1 T with the removal of a 40 MPa stress, as shown in Figs. 12 (d),  12 (e) and 12 (f) where $\Delta S$ values are 10.5 J kg$^{-1}$ K$^{-1}$ (at T = 297 K), 13.9 J kg$^{-1}$ K$^{-1}$ (at T = 298 K), and 15.1 J kg$^{-1}$ K$^{-1}$ (at $T$ = 299 K). Similar trends, are also observed for the multicaloric adiabatic temperature changes when compared to single caloric effects. The maximum multicaloric $\Delta T$= -5.9 K   obtained at $T$=298 K (Fig. 13(e))  is larger than any single caloric value (see fig.7a). Furthermore, at low magnetic fields, the values obtained by applying a 1 T field and removing a stress of 40 MPa ($\Delta T$ = - 2 K), double the magnetocaloric values obtained for 1 T ($\Delta T$= - 1 K).

It is to be noted that in general the multicaloric response of a given thermodynamic system $\Delta S(T,  0 \rightarrow \sigma, 0 \rightarrow H)$ is not given by the sum of single caloric effects $\Delta S(T,  0  \rightarrow \sigma, 0)$ and $\Delta S(T,   0, 0 \rightarrow H)$, because there is also a contribution from a cross-coupling term \cite{Planes2014,SternTaulats2018}, which accounts for the cross-response of the material to the application of non-conjugated fields. In our case, there is no contribution from such a coupling term because we are computing the multicaloric effect arising from the removal of one field (stress) and application of another field (magnetic field). In that case, $\Delta S(T, \sigma \rightarrow 0, 0 \rightarrow H)$ can be obtained  by the addition of $\Delta S(T , 0, 0 \rightarrow H )$ and $\Delta S(T, \sigma \rightarrow 0, 0, )$. 

The reversibility under field cycling is a relevant feature for a potential application of a given caloric (multicaloric) effect for refrigeration. A thorough analysis of the reversibility of the multicaloric effect in our Ni-Mn-In alloy has not been possible due to the poor quality of the thermograms recorded on cooling. However, the determined hysteresis in temperature, stress and magnetic field of the martensitic transition together with the stress and magnetic-field dependences of the transition temperatures enable us to make some estimates on the reversibility of the caloric effects. By considering a thermal hysteresis of $\sim$ 12 K, and taking representative values (see Figs. 2(b) and 2(c)) for the typical shift of the transition with field and stress of $dT/d \mu_0 H \sim$ -2 KT$^{-1}$ and $dT/d\sigma \sim$ 0.08 KMPa$^{-1}$, it is expected that the magnetocaloric effect is reversible for fields larger than 6 T, while the elastocaloric effect is expected to be reversible for stresses larger than 150 MPa. However, the effective hysteresis in the given external field can be drastically reduced by application of a secondary field \cite{Liu2012,SternTaulats2018}. While magnetic fields in the order of $\sim$ 6 T are unfeasible for practical applications, the application of stress can enhance the reversibility of the magnetocaloric effect. In our case, if we consider a magnetic field of 1 T, the magnetocaloric effect is expected to be reversible under the following sequence (which is schematized in the Supplementary Material, Fig. S4): (i) application of 1 T (in the absence of stress), which induces (partially) the transformation from martensite to austenite (ii) application of a $\sim$ 125 MPa, (iii) removal of magnetic field (keeping the stress constant), which induces the transformation from austenite to martensite and (iv) removal of the stress. Hence, it is seen that the additional application of a moderate stress ($\sigma \sim$ 100 MPa) turns Ni-Mn-In into a suitable material for refrigeration devices using permanent magnets.


\section{Conclusion}


We have used a unique  calorimeter working under magnetic field and uniaxial load to study the multicaloric response of Ni-Mn-In, a prototype metamagnetic shape-memory alloy. A numerical treatment of our calorimetric data has enabled us to obtain the entropy of the alloy over the whole temperature, magnetic-field, and uniaxial-stress phase-space. Based on this, we could compute single caloric (elastocaloric and magnetocaloric) and multicaloric effects for  arbitrary combinations of magnetic field and stress. Our results show that the multicaloric response of Ni-Mn-In exceeds that of single caloric effects. In particular, a suitable combination of magnetic field and stress gives rise to isothermal entropy and adiabatic temperature changes larger than those achievable when only a single external stimulus is applied. Furthermore, the combination of two external stimuli enlarges the temperature window where the alloy exhibits a giant caloric response, which is also accompanied by an enhancement in the reversibility of the caloric response when the external stimuli are cyclically changed.

The advantages of the multicaloric effect in relation to single caloric effects found here may  lead to clear improvements in the use of multicaloric materials for cooling applications in refrigeration devices designed to work at low values of the external fields. For instance, the combination of  a magnetic field of 1 T, with a uniaxial stress of 40 MPa, yields an isothermal entropy change of 15 J kg$^{-1}$K$^{-1}$, which is more than double the maximum value achievable for the pure magnetocaloric effect.

It is expected that many of the trends found here for a metamagnetic shape-memory alloy may also be valid for other multiferroic materials with strong coupling between different degrees of freedom. Present results should inspire the development of refrigeration devices, which take advantage of the multicaloric response of multiferroic materials.

\section{Experimental Section}

A sample with nominal composition Ni$_{50}$Mn$_{35.5}$In$_{14.5}$ was prepared by arc melting. The ingot was turned upside down and remelted several times to ensure chemical homogeneity. The button was further treated using the suction-casting option of the arc melter. The specimen was subsequently annealed at 900 $^o$C for 24 h, followed by water quenching. From the heat-treated rod, a block with dimensions 2 $\times$ 2 $\times$ 4.9 mm$^{3}$ was cut and polished. A smaller piece was cut for thermomagnetization measurements which were carried out using a Vibrating Sample Magnetometer.

Differential Scanning Calorimetry measurements were conducted by means of a bespoke differential scanning calorimeter, which enables simultaneous measurements of the length of the specimen. The system operates under magnetic fields up to 6 T, and uniaxial loads up to 1.2 kN. The calorimeter is a 15 mm diameter and 45 mm length copper block with Peltier modules on the top and bottom surfaces which are differentially connected.  The sample is placed on  a high strength aluminium disk on top of the upper surface of the top Peltier element while a dummy sample is placed on the surface of the second Peltier element. The temperature of the calorimeter is controlled by means of a cryofluid which circulates through an aluminium container that surrounds the calorimeter. 
The ensemble of the calorimeter and the aluminium container is inserted into the bore of a cryogenic free magnet. Uniaxial load is applied to the sample through a high-strength aluminium rod whose upper end is in mechanical contact with a free mobile platform and the lower end pushes an polyether ether ketone (PEEK) disk placed on top of the upper surface of the sample. Length changes are measured by means of a linear variable differential transformer (LVDT) sensor in contact with the free mobile platform that measures its relative positon. The load is applied by placing pre-weighted lead ingots on top of the mobile platform. Full details and a scheme of this device can be found in \cite{GraciaCondal2018a}.

Specific heat was measured for temperatures up to 310 K using a bespoke experimental system \cite{Stotter 2019}, for fields up to 2 T; and for temperatures up to 380 K and fields up to 4 T, using a commecial PPMS (Quantum Design) calorimeter. Temperature changes resulting from the application and removal of magnetic fields under applied uniaxial compressive load were measured by means of a purpose-built experimental system. The specimen temperature was measured by a fine gauge K thermocouple (0.075 mm diameter) in contact with the surface of the sample. The magnetic field and uniaxial load ranges are 0 - 2T and 0 - 1kN, respectively. The set-up is  described in detail in Ref. [\onlinecite{GraciaCondal2018b}].

\section*{Supplementary Material}

Supplementary Material shows a comparison between direct and indirect magnetocaloric adiabatic temperature changes. It also shows three dimensional plots for the isothermal entropy and adiabatic temperature changes for elastocaloric effects under magnetic field and magnetocaloric effects under uniaxial stress. It finally provides a scheme for a reversible multicaloric cycle.

\section*{Acknowledgements}
This work was supported by funding from CICyT (Spain), project MAT2016-75823-R, the Helmholtz Association via the Helmholtz-RSF Joint Research Group with Project No. HRSF-0045, the HLD at HZDR, a member of the European Magnetic Field Laboratory (EMFL), and the European Research Council (ERC) under the European Union's Horizon 2020 research and innovation programme (grant no. 743116-project Cool Innov). A.G. acknowledges financial support from Universitat de Barcelona under the APIF scholarship. 

\section*{Data Availability}
Data available on request from the authors.

\newpage

\section*{Supplementary}

\subsection{Magnetocaloric temperature changes.}

\begin{figure}[h]
\epsfig{file=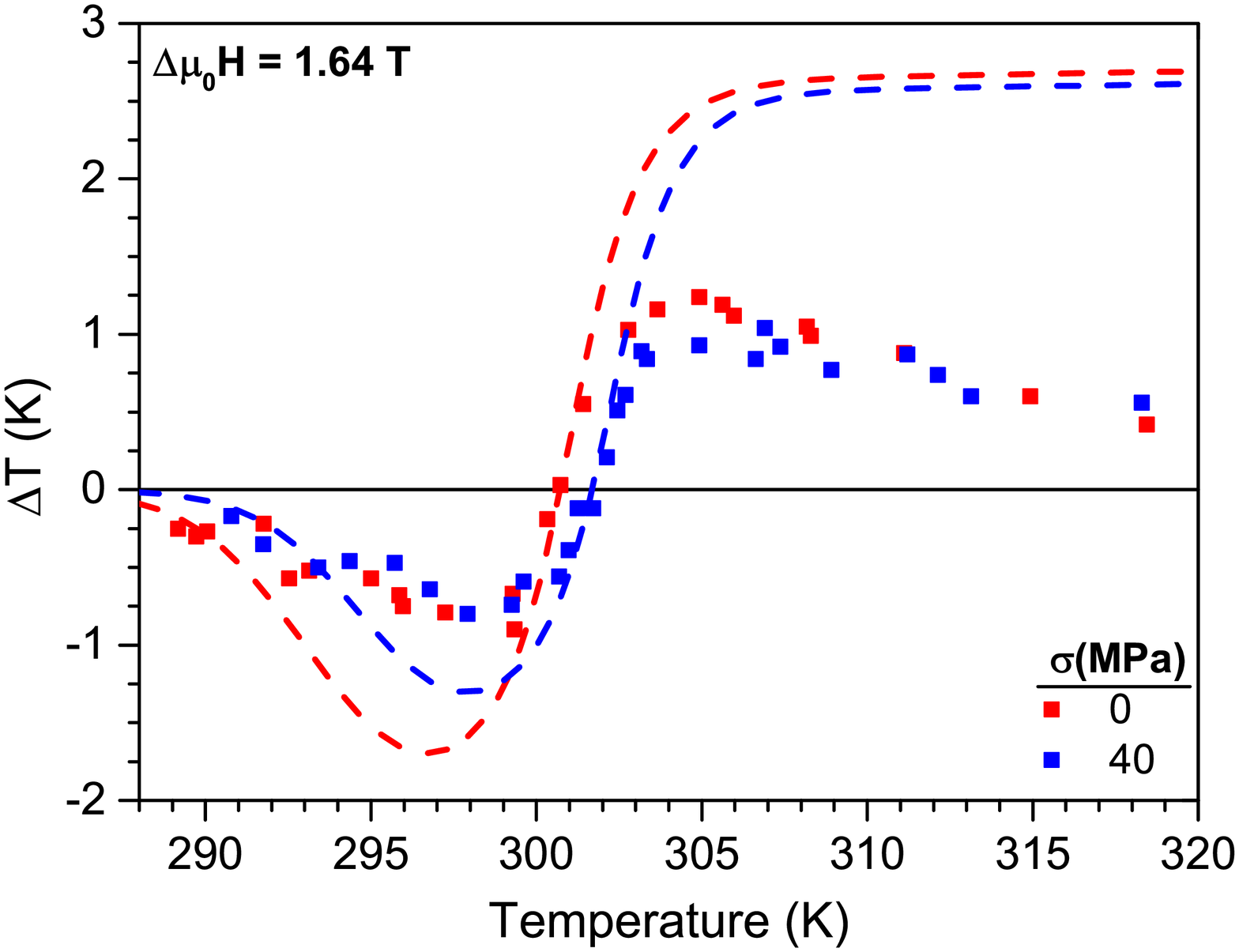,width=9cm,clip=}
\captionsetup[figure]{labelformat=empty}

{FIG. S1. Magnetocaloric adiabatic temperature changes resulting from the application of a \hspace{4mm} 1.64 T magnetic field in the absence of applied stress (red symbols and line), and under an applied uniaxial stress of 40 MPa (blue symbols and line). Symbols correspond to directly measured values and lines, to values derived form entropy curves.}
\label{figS1}
\end{figure}

Direct measurements of the adiabatic temperature change have been made using an experimental system \cite{GraciaCondal2018} which enables simultaneous application of uniaxial load and magnetic field, while the sample temperature is measured by a fine gauge K thermocouple (75$\mu$m diameter) in thermal contact with the surface of the sample. Results for a magnetic field ot 1.64 T in the absence of stress and for an applied stress (constant) of 40 MPa are shown in Figure S1 as solid symbols, and they are compared to data derived from entropy curves (lines). Direct $\Delta T$ data confirm the cross-over form the  inverse to the conventinal magnetocaloric effects. The temperature region where the inverse magnetocaloric effect derived from entropy curves takes place perfectly matches the region determined from direct measurements. For the inverse magnetocaloric effect, the differences between measured and computed  $\Delta T$ values ($\sim$ 0.7 K) are mostly due to the lack of absolute adiabaticity of the thermometric system, and are in the order of those expected for that particular device \cite{SternTaulats2015}. In contrast, larger differences are found for the conventional magnetocaloric effect. In that case, the discrepancy between measured and computed data must be ascribed to the fact that quasi-direct methods provide reliable $\Delta T$ values around a first-order phase transition, however, $\Delta T$ data at temperatures beyond the phase transtion can be affected by a considerable error.



\subsection{Single Caloric  entropy and temperature changes.}

\begin{figure}[h]
\epsfig{file=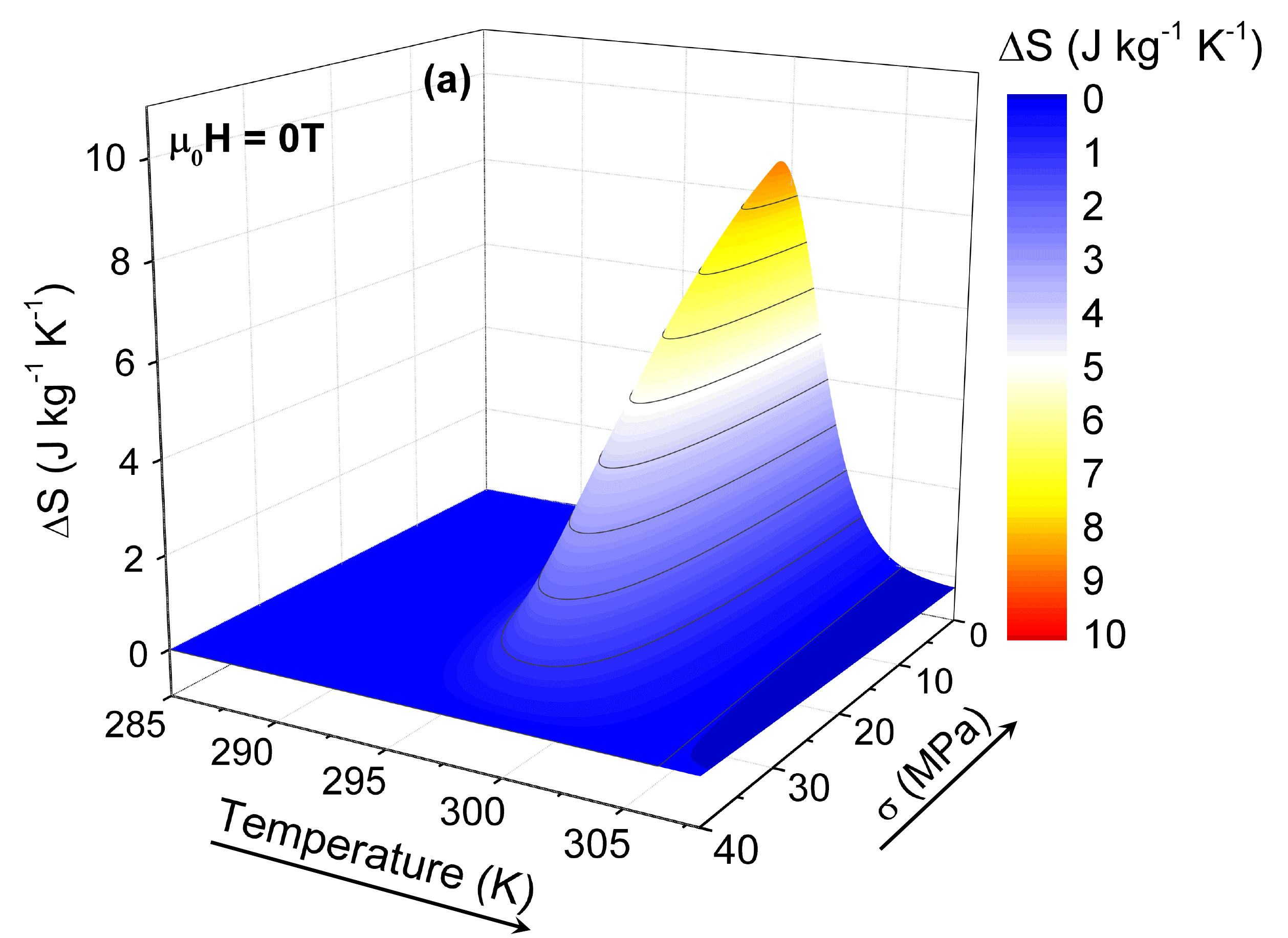,width=8cm,clip=}
\epsfig{file=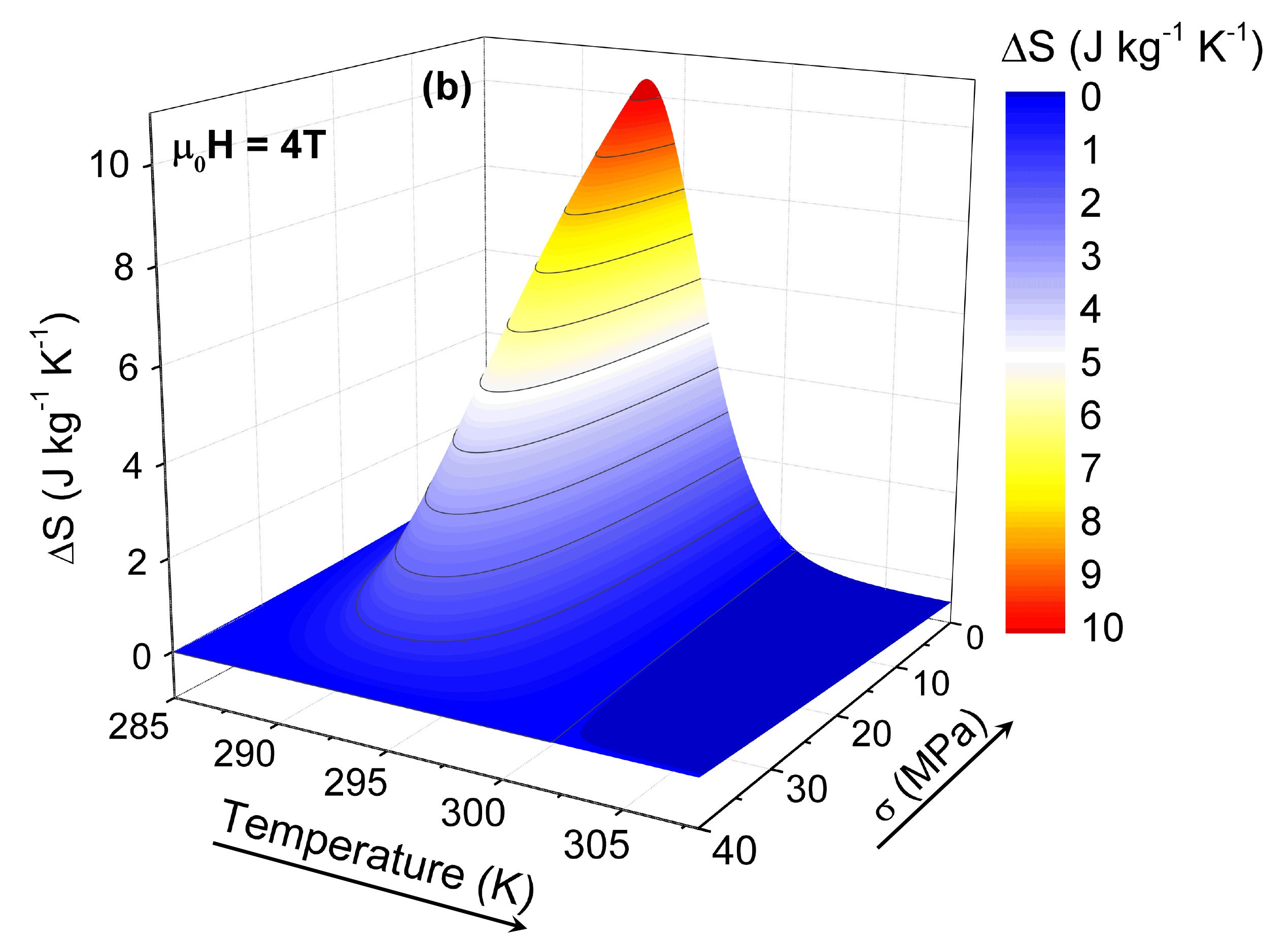,width=8cm,clip=}
\epsfig{file=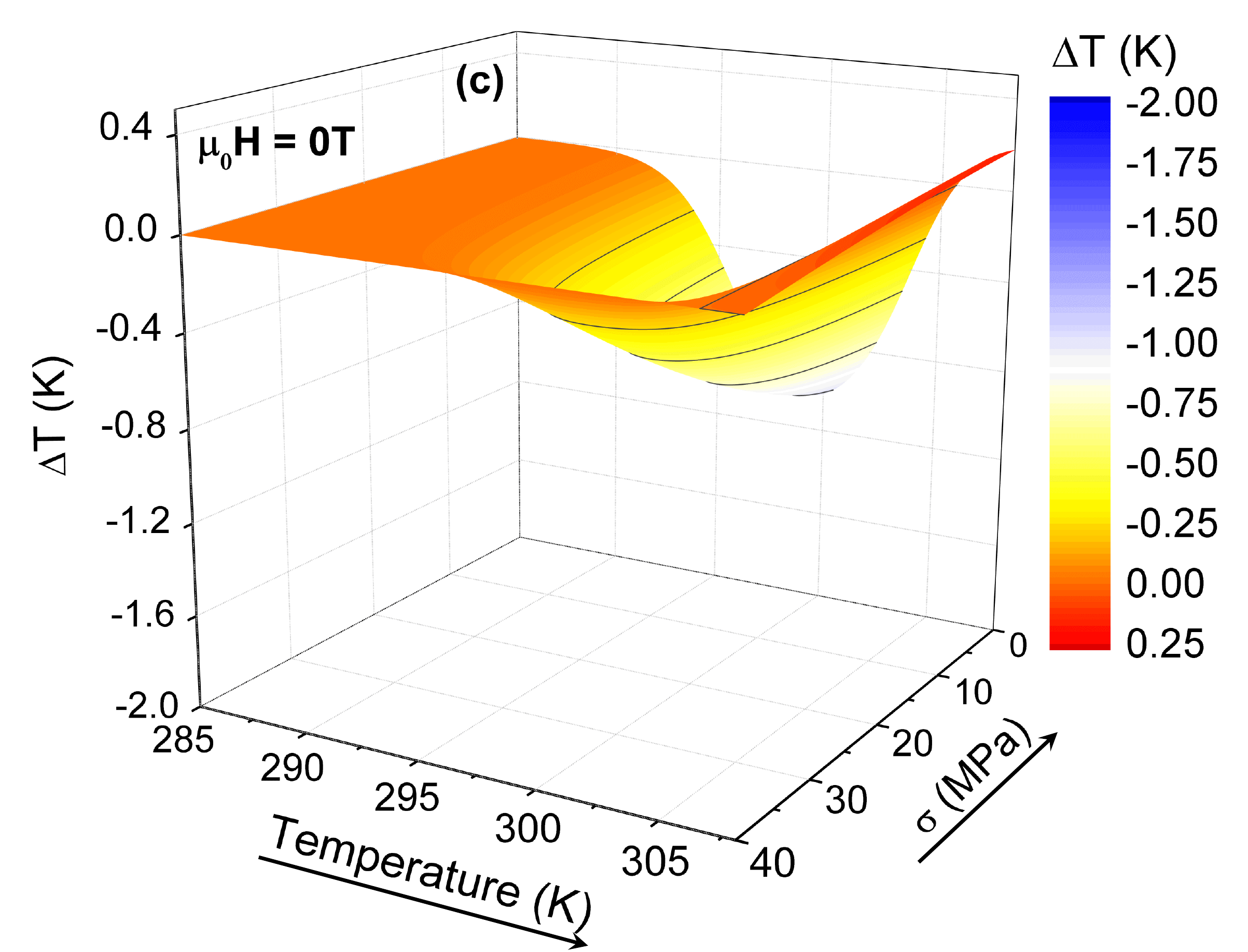,width=8cm,clip=}
\epsfig{file=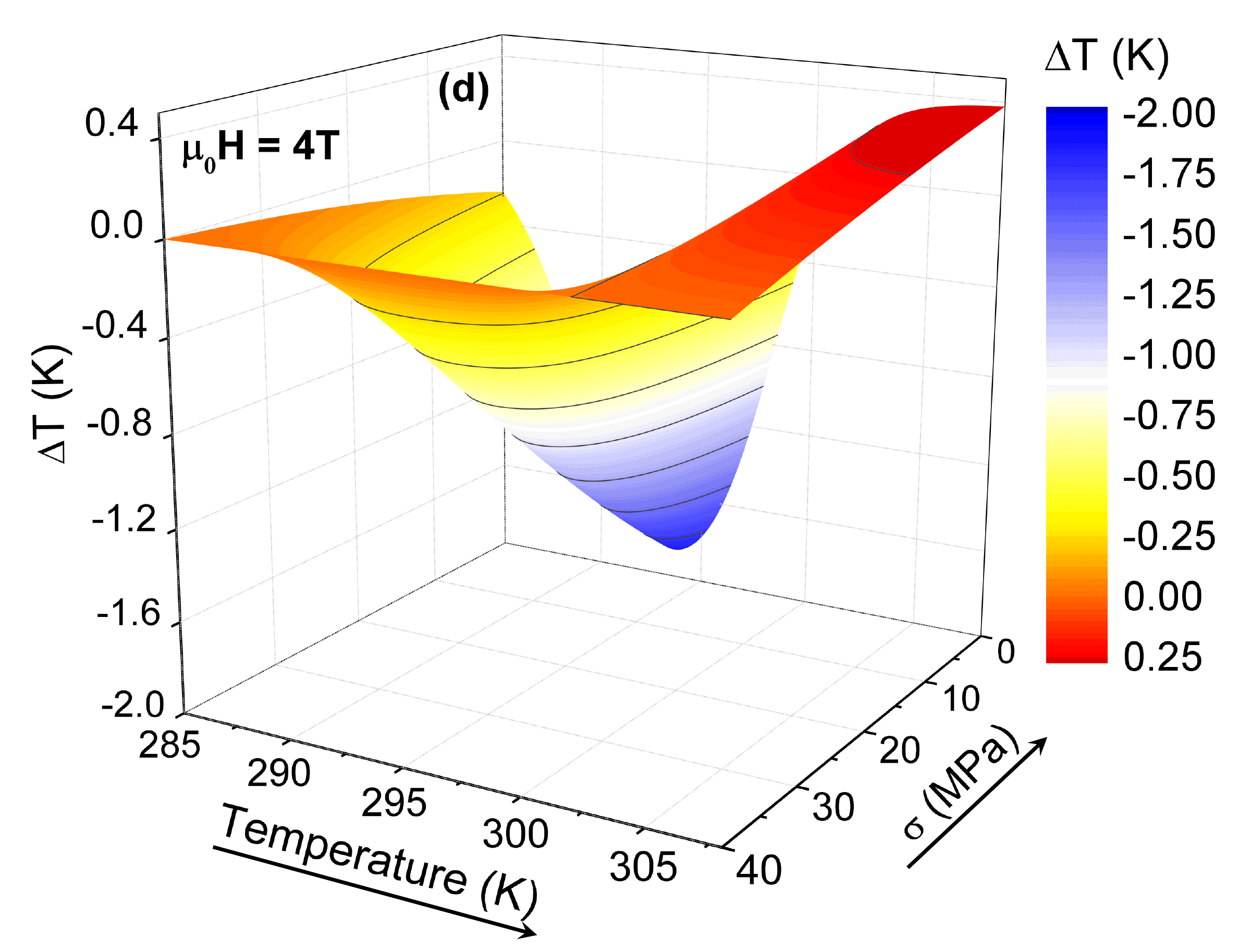,width=8cm,clip=}

{FIG. S2. Elastocaloric effect. (a) and (b) Isothermal entropy changes, and (c) and (d)  adiabatic temperature changes. Left panels (a) and (c) correspond to the absence of magnetic field, and right panels (b) and (d), under an applied (constant) magnetic field of 4 T. Data corresponds to the removal of a uniaxial stress from an initial value of 40 MPa to an arbitrary value $\sigma$, as indicated by the arrows in the stress axis.} 
\label{figS2}
\end{figure}


\begin{figure}[h]
\epsfig{file=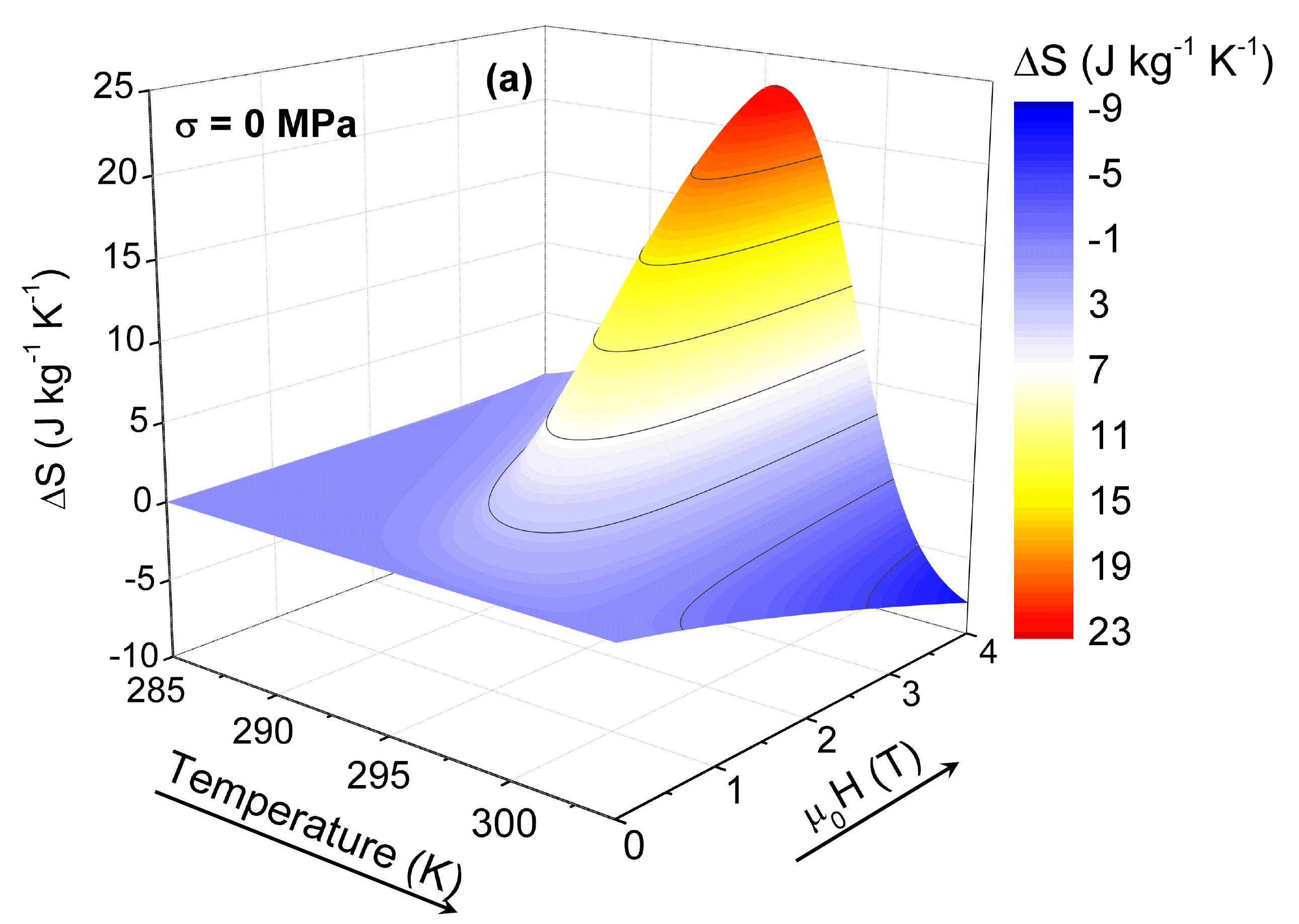,width=8cm,clip=}
\epsfig{file=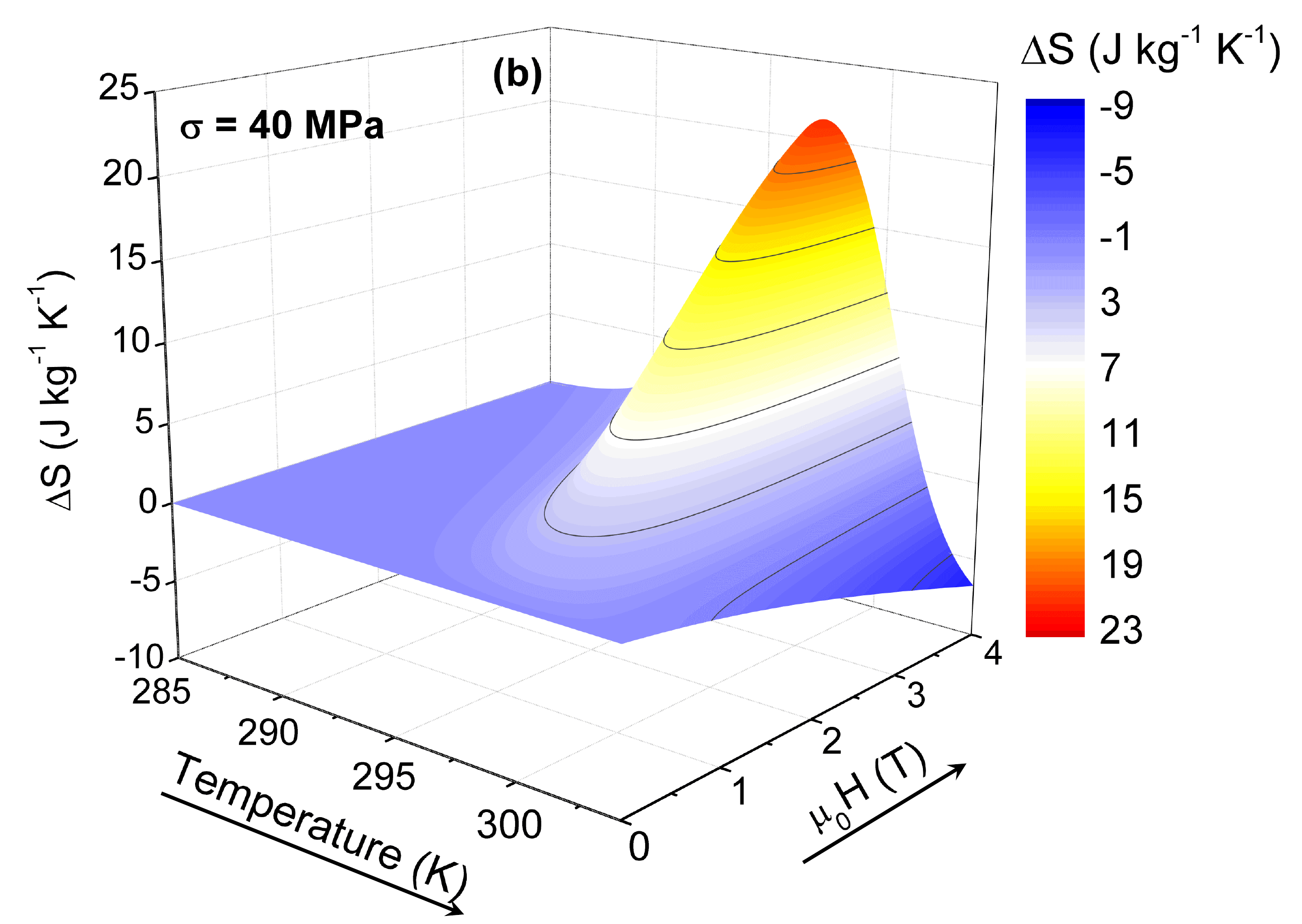,width=8cm,clip=}
\epsfig{file=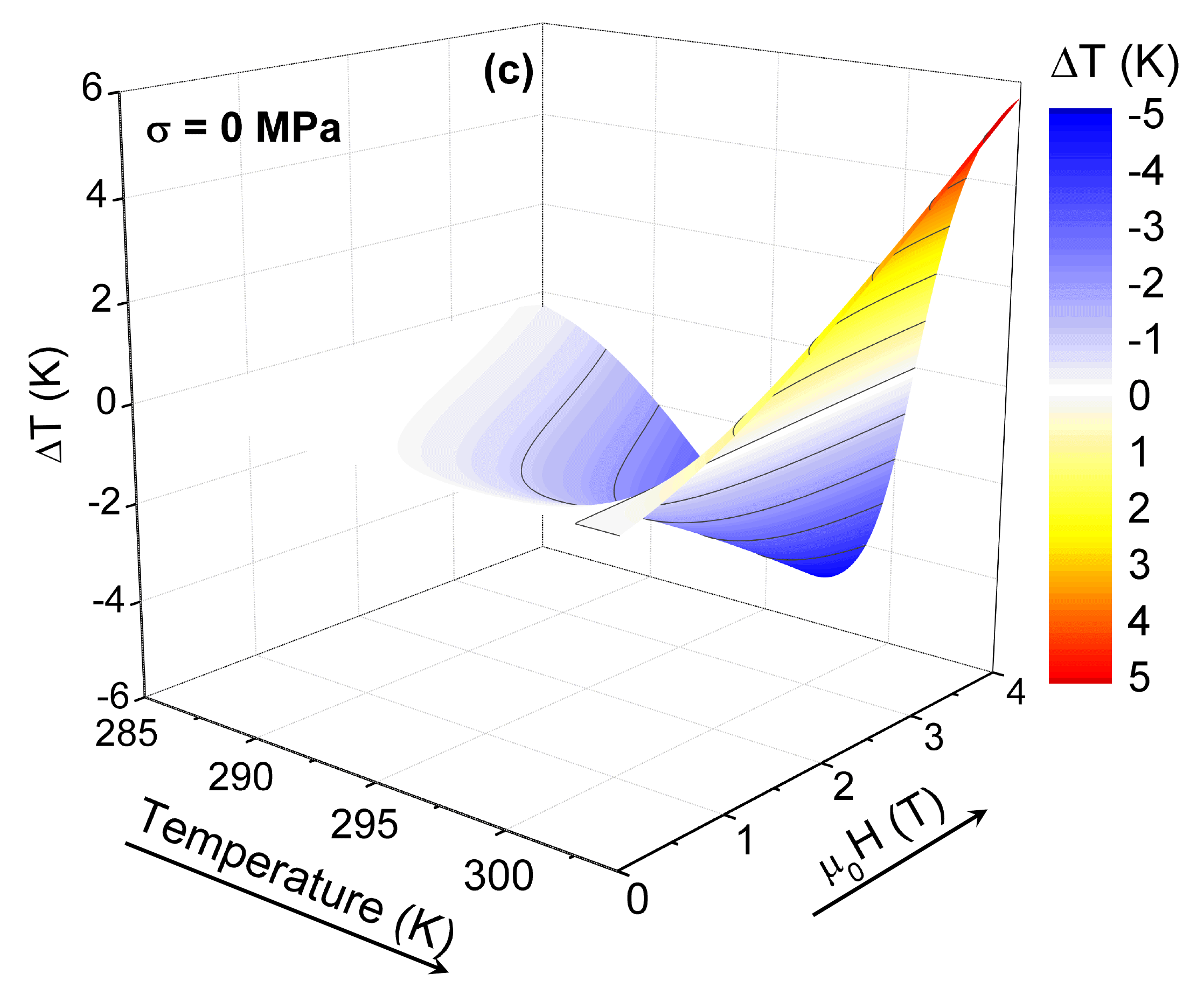,width=8cm,clip=}
\epsfig{file=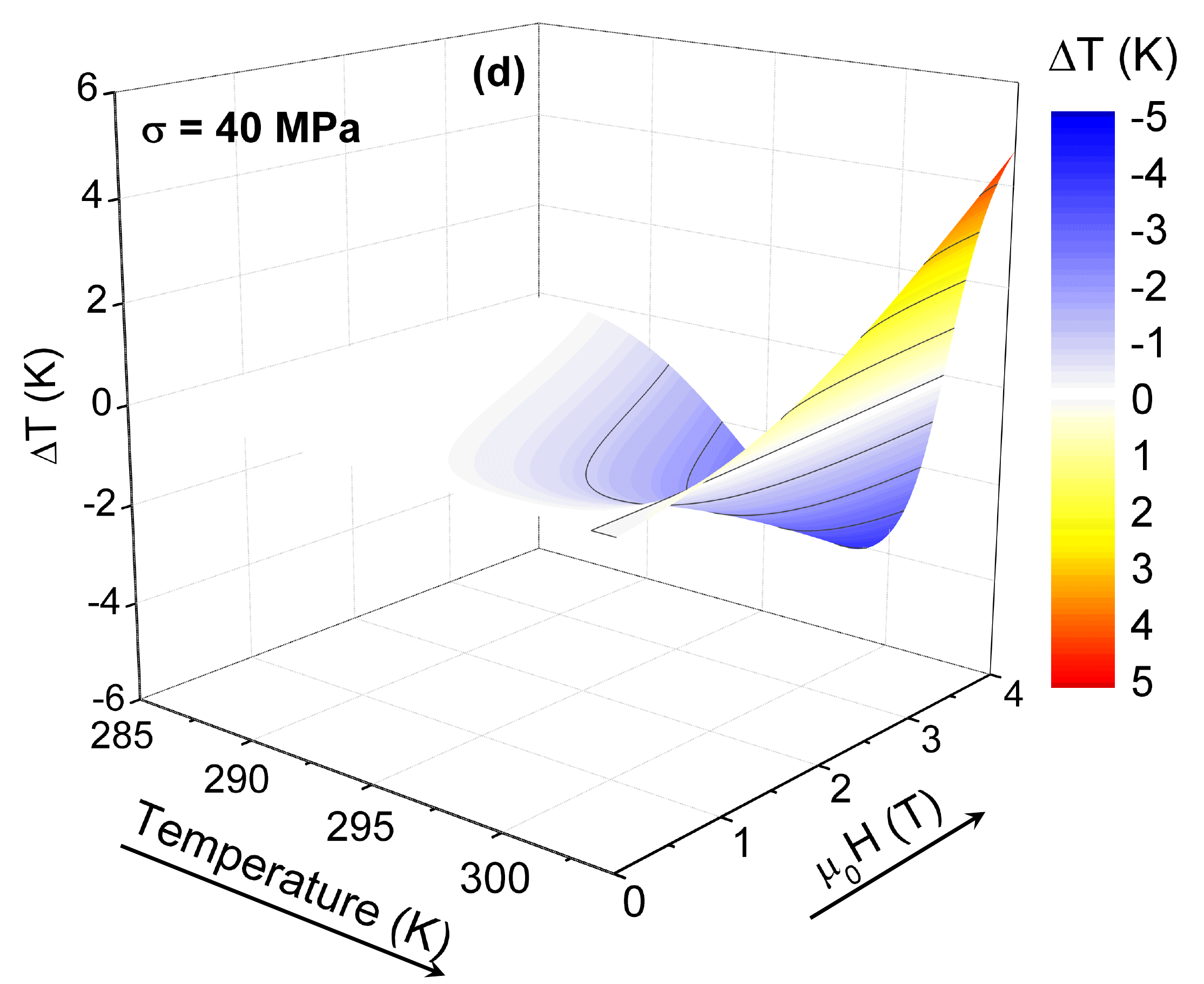,width=8cm,clip=}
\captionsetup[figure]{labelformat=empty}

{FIG. S3. Magnetocaloric effect. (a) and (b) Isothermal entropy changes, and (c) and (d) adiabatic temperature changes. Left panels (a) and (c) correspond to the absence of uniaxial stress, and right panels (b) and (d), under an applied uniaxial stress of 40 MPa. Data corresponds to the application of a magnetic field from zero to an arbitrary value $\mu_0 H$, as indicated by the arrows in the field axis. \vspace{6cm}}
\label{figS3}
\end{figure}

\newpage

\subsection{Reversible multicaloric cycle}

\begin{figure}[h]
\epsfig{file=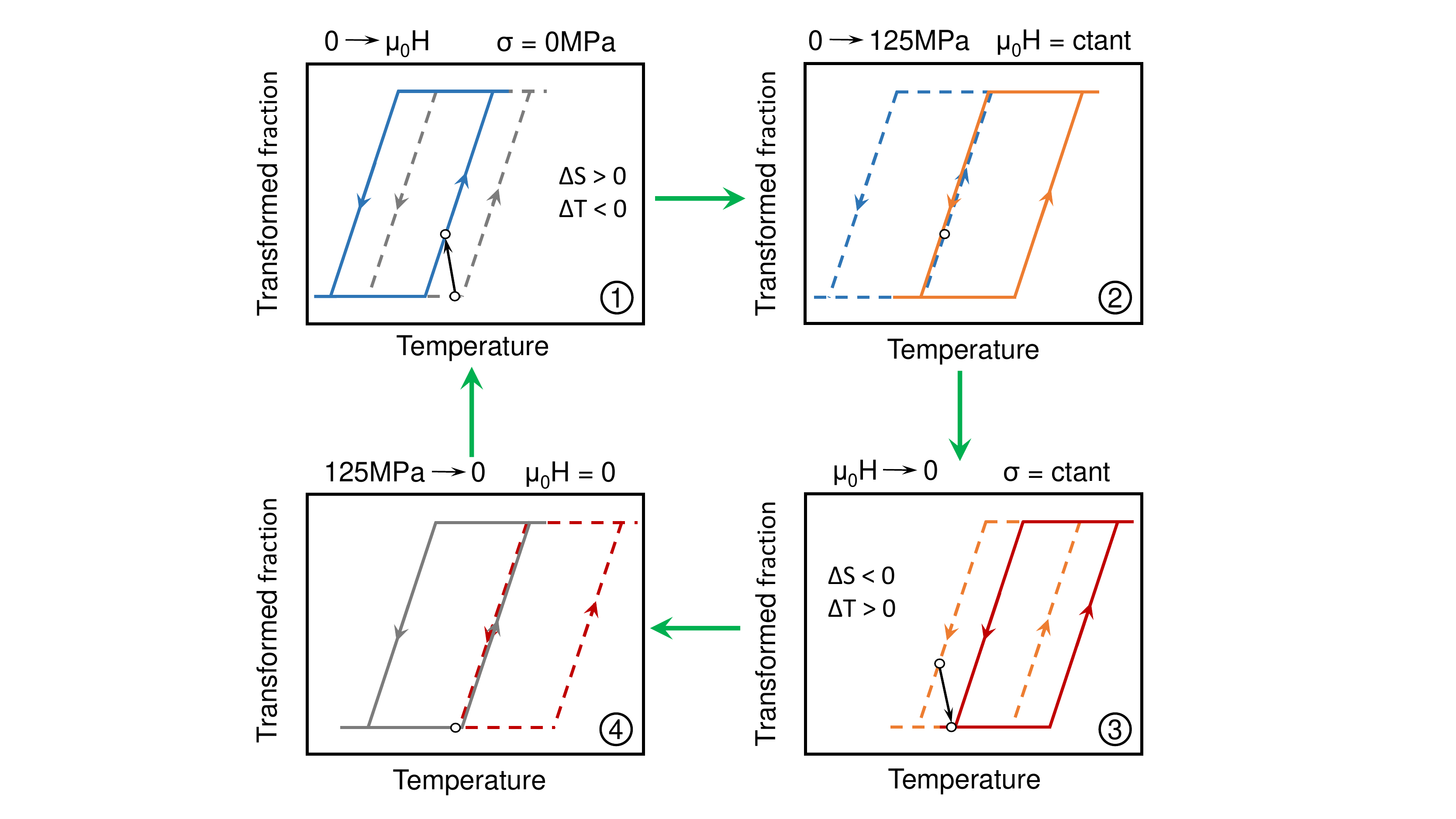,width=15cm,clip=}
\captionsetup[figure]{labelformat=empty}

{FIG. S4. Sketch of a multicaloric reversible cycle showing the fraction of austenite as a function of temperature, for selected values of magnetic field and uniaxial stress. Panel 1: Application of magnetic field in the absence of stress. Panel 2: Application of stress under an applied (constant) magnetic field. Panel 3: Removal of magnetic field under an applied (constant) stress. Panel 4: Removal of stress in the absence of magnetic field.}
\label{figS4}
\end{figure}

Figure S4 sketches a multicaloric reversible cycle under the combined action of magnetic field and uniaxial stress (for the sake of simplicity we do not consider partial hysteresis loops). Panel 1: application of a 1 T magnetic field shifts the hysteresis loop to lower temperatures (blue loop), and the sample (initially in the martensitic phase) partially transforms to austenite (indicated by the black arrow). Panel 2: Application of 125 MPa stress shifts the hysteresis loop to higher temperatures (orange loop) so that  the state of the sample (under magnetic field and stress) lies on the decreasing branch of the hysteresis loop. Panel 3: While keeping the stress constant, removal of magnetic field shifts the hysteresis loop to higher temperatures (red loop) and the sample transforms to the martensitic phase (black arrow). Panel 4: Removal of stress shifts the hysteresis loop to lower temperatures (grey loop) and brings the sample to its initial state.




\begin{thebibliography}{1}
\bibitem{footnote1} The spread in temperature of the martensitic transtion imply that the surfaces separating martensite from austenite are not truly two-dimensional (planes) but rather they have a certain thickness. This effect is not considered here.
\end{thebibliography}
\end{document}